\documentclass{article}%
\usepackage{geometry}
\usepackage{amsmath}
\usepackage{amsfonts}
\usepackage{amssymb}
\usepackage{cancel}
\usepackage{enumitem}
\usepackage{graphicx}
\usepackage{epstopdf}
\usepackage{color}
\usepackage{soul}%
\usepackage{cite}
\usepackage{float}
\usepackage{booktabs}
\usepackage{enumitem}
\usepackage{array}
\providecommand{\U}[1]{\protect\rule{.1in}{.1in}}

\newcolumntype{M}{>{\centering\arraybackslash}m{\dimexpr.25\linewidth-2\tabcolsep}}
\DeclareMathOperator\erf{erf}

\begin{document}

\title{\textbf{A moment-equation-copula-closure method for nonlinear vibrational systems subjected to correlated noise}}
\author{Han Kyul Joo and Themistoklis P.\ Sapsis\thanks{Corresponding author: sapsis@mit.edu, Tel: +1-617-324-7508,  Fax: +1-617-253-8689.}\\Department of Mechanical Engineering, Massachusetts Institute of Technology,\\
77 Massachusetts Ave., Cambridge, Massachusetts 02139, USA}
\date{\today}
\maketitle
\begin{abstract}
We develop a moment-equation-copula-closure method for the inexpensive approximation of the steady state statistical structure of strongly nonlinear systems which are subjected to correlated excitations. Our approach relies on the derivation of moment equations that describe the dynamics governing the two-time statistics. These are combined with a non-Gaussian pdf representation for the joint response-excitation statistics, based on copula functions that has i) single time statistical structure consistent with the analytical solutions of the Fokker-Planck equation, and ii) two-time statistical structure with Gaussian characteristics. Through the adopted pdf representation, we derive a closure scheme which we formulate in terms of a consistency condition involving the second order statistics of the response, the \textit{closure constraint}. A similar condition, the  \textit{dynamics constraint}, is also derived directly through the moment equations. These two constraints are formulated as a low-dimensional minimization problem with respect to the unknown parameters of the representation, the minimization of which imposes an interplay between the dynamics and the adopted closure. The new method allows for the semi-analytical representation of the two-time, non-Gaussian structure of the solution as well as the joint statistical structure of the response-excitation over different time instants. We demonstrate its effectiveness through the application on bistable nonlinear single-degree-of-freedom energy harvesters with mechanical and electromagnetic damping, and we show that  the results compare favorably with direct Monte-Carlo simulations. 
\end{abstract}


\section{\label{sec_intro} Introduction}
In numerous systems in engineering, uncertainty in the dynamics is as important as the known conservation laws. Such an uncertainty can be introduced by external stochastic excitations, e.g. energy harvesters or structural systems subjected to ocean waves, wind excitations, earthquakes, and impact loads  \cite{Grigoriu02, Stratonovich67, sobczyk01, Soong93, Naess12, To11}. For these cases, deterministic models cannot capture or even describe the essential features of the response and to this end, understanding of the system dynamics and optimization of its parameters for the desired performance is a challenging task. On the other hand, a probabilistic perspective can, in principle, provide such information but then the challenge is the numerical treatment of the resulted descriptive equations, which are normally associated with prohibitive computational cost. 

The focal point of this work is the development of a semi-analytical method for the inexpensive probabilistic description of nonlinear vibrational systems of low to moderate dimensionality subjected to correlated inputs. Depending on the system dimensionality and its dynamical characteristics, numerous techniques have been developed to quantify the response statistics, i.e. the probability density function (pdf) for the system state. {For systems subjected to white noise, Fokker-Planck-Kolmogorov (FPK) equation provides a complete statistical description of the response statistics \cite{Wojtkiewicz99, Dunne97, Paola02}. However, exact analytical solutions of the FPK equation are available only for a small class of systems. An alternative computational approach, the path integral solution (PIS)  method, has  been developed to provide the response pdf for general nonlinear systems at a specific time instant given  the pdf of an earlier time instant. Many studies have been focused on the application of step-by-step PIS method numerically \cite{Wehner83, Naess93, Paola08} and analytically \cite{Kougioumtzoglou12, Kougioumtzoglou14, Matteo14} reporting its effectiveness on capturing the response statistics.} On the other hand, for non-Markovian systems subjected to correlated excitations the joint response-excitation pdf method provides a computational framework for the full statistical solution \cite{Sapsis08, Venturi12, Cho13}. However, such methodologies rely on the solution of transport equations for the pdf and they are associated with very high computational cost especially when it comes to the optimization of system parameters. 

To avoid solving the transport equations for the pdf, semi-analytical approximative approaches with significantly reduced computational cost have been developed. Among them the most popular method in the context of structural systems is the statistical linearization method \cite{Caughey59, Caughey63, Kazakov54, Roberts03, Socha08}, which can also handle correlated excitations. The basic concept of this approach is to replace the original nonlinear equation of motion with a linear equation, which can be treated analytically, by minimizing the statistical difference between those two equations. Statistical linearization performs very well for systems with unimodal statistics, i.e. close to Gaussian. However, when the response is essentially nonlinear, e.g. as it is the case for a double-well oscillator, the application of statistical linearization is less straightforward and involves the ad-hoc selection of shape parameters for the response statistics \cite{Crandall05}. 

An alternative class of methods relies on the derivation of moment equations, which describes the evolution of the the joint response-excitation statistical moments or (depending on the nature of the stochastic excitation) the response statistical moments \cite{Sancho70, Bover78, Beran94}. The challenge with moment equations arises if the equation of motion of the system contains nonlinear terms in which case we have the well known closure problem. This requires the adoption of closure schemes, which essentially truncate the infinite system of moment equations to a finite one. The simplest approach along this line is the Gaussian closure \cite{Iyengar78} but nonlinear closure schemes have also been developed (see e.g. \cite{Crandall80, Crandall85, Liu88, Wu84, Ibrahim85, Grigoriu91, Hasofer95, Wojtkiewicz96, Grigoriu99}). In most cases, these nonlinear approaches may offer some improvement compared with the stochastic linearization approach applied to nonlinear systems but the associated computational cost is considerably larger \cite{Noori87}. For strongly nonlinear systems, such as bistable systems, these improvements can be very small. Bistable systems, whose potential functions have bimodal shapes, have become very popular in energy harvesting applications \cite{Green12, harne13, Daqaq11, Halvorsen13, Green13, He14, Mann09, Barton10}, where there is a need for fast and reliable calculations that will be able to resolve the underlying nonlinear dynamics in order to provide with optimal parameters of operation (see e.g. \cite{Joo14, Kluger15}).

The goal of this work is the development of a closure methodology that can overcome the limitations of traditional closure schemes and can approximate the steady state statistical structure of bistable systems excited by correlated noise. We first formulate the moment equations for the joint pdf of the response and the excitation at two arbitrary time instants \cite{Athanassoulis13}. To close the resulted system of moment equations, we formulate a two-time representation of the joint response-excitation pdf using copula functions. We choose the representation so that the single time statistics are consistent in form with the Fokker-Planck-Kolmogorov solution in steady state, while the joint statistical structure between two different time instants is represented with a Gaussian copula density. Based on these two ingredients (dynamical information expressed as moment equations and assumed form of the response statistics), we formulate a minimization problem with respect to the unknown parameters of the pdf representation so that both the moment equations and the closure induced by the representation are optimally satisfied. For the case of unimodal systems, the described approach reproduces the statistical linearization method while for bi-modal systems it still provides meaningful and accurate results with very low computational cost. 

The developed approach allows for the inexpensive and accurate approximation of the second order statistics of the system even for oscillators associated with double-well potentials. In addition, it allows for the semi-analytical approximation of the full non-Gaussian joint response-excitation pdf in a post-processing manner. We illustrate the developed approach through nonlinear single-degree-of-freedom energy harvesters with double-well potentials subjected to correlated noise with Pierson-Moskowitz power spectral density. We also consider the case of bi-stable oscillators coupled with electromechanical energy harvesters (one and a half degrees-of-freedom systems), and we demonstrate how the proposed probabilistic framework can be used for performance optimization and parameters selection.

\section{Description of the Method}
In this section, we give a detailed description of the proposed method for the inexpensive computation of the response statistics for dynamical systems subjected to colored noise excitation. The computational approach relies on two basic ingredients:
\begin{itemize}
        \item \textit{Two-time statistical moment equations}. These equations will be derived directly from the system equation and they will express the dynamics that govern the two-time statistics. For systems excited by white-noise, single time statistics are sufficient to describe the response but for correlated excitation, this is not the case and it is essential to consider higher order moments. Note that higher (than two) order statistical moment equations may be used but in the context of this work two-time statistics would be sufficient.
        \item \textit{Probability density function (pdf) representation for the joint response-excitation statistics}. This will be a family of probability density functions with embedded statistical properties such as multi-modality, tail decay properties, correlation structure between response and excitation, or others. The joint statistical structure will be represented using copula functions. We will use representations inspired by the analytical solutions of the dynamical system when this is excited by white noise. These representations will reflect features of the Hamiltonian structure of the system and will be used to derive appropriate closure schemes that will be combined with the moment equations.
\end{itemize}   
Based on these two ingredients, we will formulate a minimization problem with respect to the unknown parameters of the pdf representation so that both the moment equations and the closure induced by the representation are optimally satisfied. We will see that for the case of unimodal systems the described approach reproduces the statistical linearization method while for bi-modal systems it still provides meaningful and accurate results with very low computational cost. 

For the sake of simplicity, we will present our method through a specific system involving a nonlinear SDOF oscillator with a double well potential. This system has been studied extensively in the context of energy harvesting especially for the case of white noise excitation {\cite{Daqaq11, Daqaq12, Gammaitoni09, Ferrari10}}. However, for realistic setups it is important to be able to optimize/predict its statistical properties under general (colored) excitation. More specifically we consider a nonlinear harvester of the form 
\begin{align}
\ddot{x} + {\lambda}\dot{x}+{k}_1x+{k}_3x^3=\ddot{y}.
\label{GE}
\end{align}
where $x$ is the relative displacement between the harvester mass and the base, $y$ is  the base excitation representing a stationary stochastic process, ${\lambda}$ is normalized (with respect to mass) damping coefficient, and ${k}_1$ and ${k}_3$ are normalized stiffness coefficients.

\begin{figure}[H]
\centerline{
\begin{minipage}{\hsize}\begin{center}
\includegraphics[width=0.35\hsize]{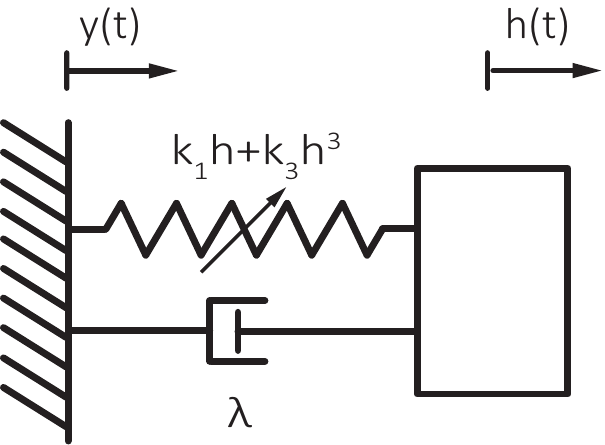}
\end{center}\end{minipage}
}
\caption{Nonlinear energy harvester with normalized system parameters.}
\end{figure}

\subsection{Two-time Moment System}

\noindent
We consider two generic time instants, $t$ and $s$. The two-time moment equations have been considered previously in \cite{Athanassoulis13} for the determination of the solution of a `half' degree-of-freedom nonlinear oscillator by utilizing a Gaussian closure. We multiply the equation of motion at time $t$ with the response displacement $x(s)$ and apply the mean value operator $\overline{\square}$ (ensemble average). This will give us an equation which contains an unknown term on the right hand side. To determine this term we repeat the same step but we multiply the equation of motion with $y(s)$. This gives us the following two-time moment equations:
\begin{align}
 \overline{\ddot{x}(t)y(s)} +{\lambda} \overline{\dot{x}(t)y(s)}+{k}_1\overline{x(t) y(s)}+{k}_3\overline{x(t)^3y(s)}=  \overline{\ddot{y}(t)y(s)},\\
 \overline{\ddot{x}(t)x(s)} + {\lambda}\overline{\dot{x}(t)x(s)}+{k}_1\overline{x(t) x(s)}+{k}_3\overline{x(t)^3x(s)}=\overline{\ddot{y}(t)x(s)}.
\end{align}
Here the excitation is assumed to be a stationary stochastic process with zero mean and a given power spectral density; this can have an arbitrary form, e.g. monochromatic, colored, or white noise. Since the system is characterized by an odd restoring force, we expect that its response also has zero mean. Moreover, we assume that after an initial transient the system will be reaching a statistical steady state given the stationary character of the excitation. Based on properties of mean square calculus \cite{sobczyk01, Beran94}, we interchange the differentiation and the mean value operators. Then the moment equations will take the form:
\begin{align}
\frac{\partial^2}{\partial t^2} \overline{{x}(t)y(s)} +{\lambda} \frac{\partial}{\partial t}\overline{{x}(t)y(s)}+{k}_1\overline{x(t) y(s)}+{k}_3\overline{x(t)^3y(s)}=  \frac{\partial^2}{\partial t^2}\overline{{y}(t)y(s)},\label{MomentRaw1}\\
\frac{\partial^2}{\partial t^2} \overline{{x}(t)x(s)} + {\lambda}\frac{\partial}{\partial t} \overline{{x}(t)x(s)}+{k}_1\overline{x(t) x(s)}+{k}_3\overline{x(t)^3x(s)}= \frac{\partial^2}{\partial t^2} \overline{{y}(t)x(s)}.
\label{MomentRaw2}
\end{align}
Expressing everything in terms of the covariance functions, above equations will result in:
\begin{align}
\frac{\partial^2}{\partial t^2} C_{xy}^{ts} +{\lambda}\frac{\partial}{\partial t} C_{xy}^{ts} +{k}_1 C_{xy}^{ts} + {k}_3\overline{x(t)^3y(s)} =\frac{\partial^2}{\partial t^2} C_{yy}^{ts},\label{Moment_ts1}\\
\frac{\partial^2}{\partial t^2} C_{xx}^{ts} +{\lambda}\frac{\partial}{\partial t} C_{xx}^{ts} +{k}_1C_{xx}^{ts} + {k}_3\overline{x(t)^3x(s)}=\frac{\partial^2}{\partial t^2} C_{yx}^{ts}, \label{Moment_ts2}
\end{align}
where the covariance function is defined as
\begin{align}
C_{xy}^{ts} = \overline{{x}(t)y(s)} = C_{xy}(t-s) = C_{xy} (\tau).
\end{align}
Taking into account the assumption for a stationary response (after the system has gone through an initial transient phase), the above moment equations can be rewritten in terms of the time difference $\tau=t-s$:
\begin{align}
\frac{\partial^2}{\partial \tau^2} C_{xy}(\tau) +{\lambda}\frac{\partial}{\partial \tau} C_{xy}(\tau) +{k}_1 C_{xy}(\tau)  + {k}_3\overline{x(t)^3y(s)} =&\ \frac{\partial^2}{\partial \tau^2} C_{yy}(\tau),\label{mt_tau1}\\
\frac{\partial^2}{\partial \tau^2} C_{xx}(\tau) +{\lambda}\frac{\partial}{\partial \tau} C_{xx}(\tau) +{k}_1C_{xx}(\tau)  + {k}_3\overline{x(t)^3x(s)} =&\ {\frac{\partial^2}{\partial \tau^2} C_{xy}(-\tau)}.\label{mt_tau2}
\end{align}
Note that all the linear terms in the original equation of motion are expressed in terms of covariance functions, while the nonlinear (cubic) terms show up in the form of fourth order moments. To compute the latter we will need to adopt an appropriate closure scheme.

\subsection{Two-time PDF  representations and induced closures}
In the absence of higher-than-two order moments, the response statistics can be analytically obtained in a straightforward manner. However, for higher order terms it is necessary to adopt an appropriate closure scheme that  closes the infinite system of moment equations. A standard approach in this case, which performs very well for unimodal systems, is the application of Gaussian closure which utilizes Isserlis' Theorem \cite{Isserlis18} to connect the higher order moments with the second order statistical quantities. Despite its success for unimodal systems, Gaussian closure does not provide accurate results for bistable systems. This is because in this case (i.e. bistable oscillators) the closure induced by the Gaussian assumption does not reflect the properties of the system attractor in the statistical steady state.

Here we aim to solve this problem by proposing a non-Gaussian representation for the joint response-response pdf at two different time instants and for the joint response-excitation pdf at two different time instants. These representations will:
\begin{itemize}
 \item incorporate specific properties or information about the response pdf (single time statistics) in the statistical steady state,
 \item capture the correlation structure between the statistics of the response and/or excitation at different time instants by employing Gaussian copula density functions,
 \item have a consistent marginal with the excitation pdf (for the case of the joint response-excitation pdf).

\end{itemize} 

\subsubsection{Representation Properties for Single Time Statistics}
\begin{figure}[htbp]
\centerline{
\begin{minipage}{\hsize}\begin{center}
\includegraphics[width=\hsize]{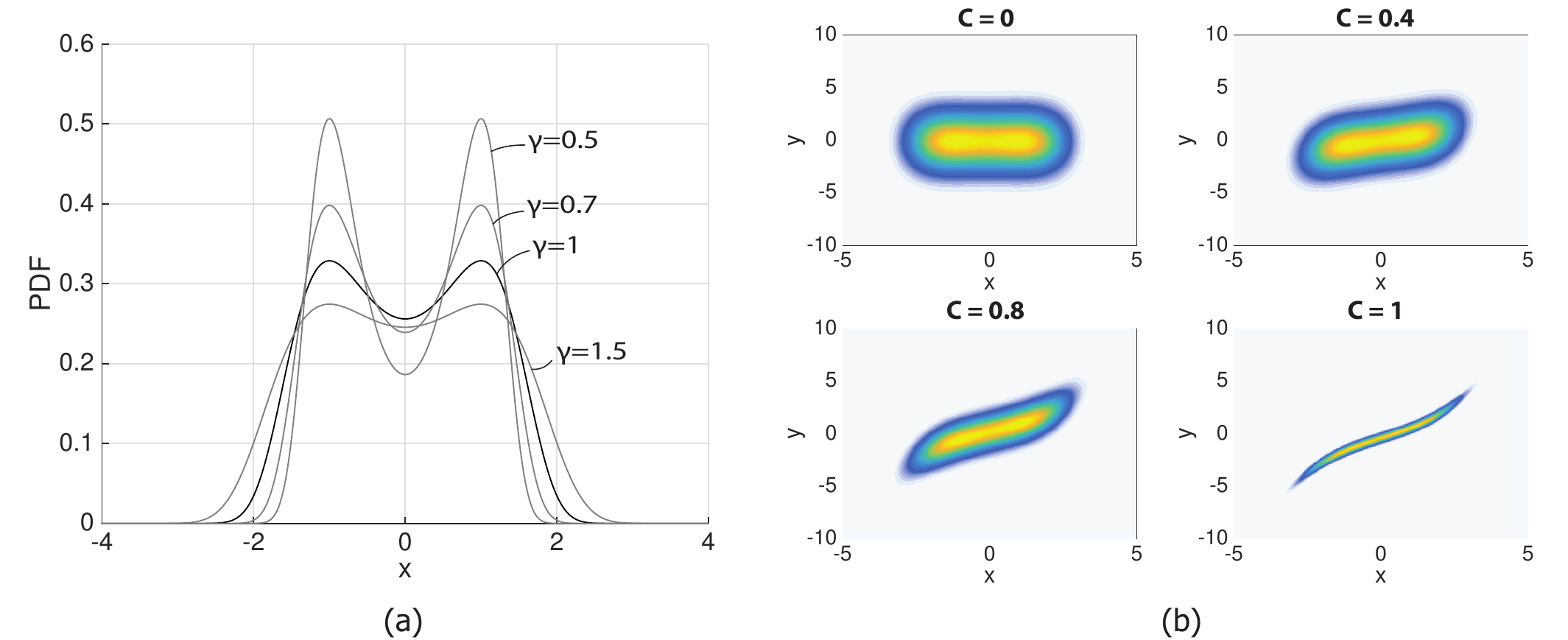}
\end{center}\end{minipage}
}
\caption{(a) Representation of the steady state pdf for single time statistics of a system with double-well potential. The pdf is shown for different energy levels of the system. (b) The joint response excitation pdf is also shown for different values of the correlation parameter $c$ ranging from small values (corresponding to large values of $|\tau|$) to larger ones (associated with smaller values of $|\tau|$).}
\label{figure2}
\end{figure}

We begin by introducing the pdf properties for the single time statistics. The selected representation will be based on the analytical solutions of the Fokker-Planck equation which are available for the case of white noise excitation \cite{Soize94, sobczyk01}, and for vibrational systems that has an underlying Hamiltonian structure. Here we will leave the energy level of the system as a free parameter - this will be determined later. In particular, we will consider the following family of pdf solutions (Figure \ref{figure2}a):
\begin{align}
f(x;\gamma) = \frac { 1 }{ \mathcal{ F } } \exp\{ -\frac { 1 }{ \gamma  } U(x)\}  = \frac{1}{\mathcal{F}}\exp\Big\{-\frac{1}{\gamma}\Big(\frac{1}{2}{k}_1 x^2 + \frac{1}{4}{k}_3x^4 \Big)\Big\},
\label{pdf_new}
\end{align}
where $U$ is the potential energy of the oscillator, $\gamma$ is a free parameter connected with the energy level of the system, and $\mathcal{F}$ is the normalization constant expressed as follows:
\begin{align}
\mathcal{F}=\int^{\infty}_{-\infty} \exp\Big\{-\frac{1}{\gamma}\Big(\frac{1}{2}{k}_1 x^2 +\frac{1}{4} {k}_3x^4 \Big)\Big\} dx.
\end{align}

\noindent
\subsubsection{Correlation Structure between Two-time Statistics}
Representing the single time statistics is not sufficient since for non-Markovian systems (i.e.  correlated excitation) the system dynamics can be effectively expressed only through (at least) two-time statistics. To represent the correlation between two different time instants we introduce Gaussian copula densities \cite{Nelsen07, Meyer13}. A copula is a multivariate probability distribution with uniform marginals. It has emerged as an useful tool for modeling stochastic dependencies allowing the separation of dependence modeling from the given marginals \cite{Qu12}. Based on this formulation we obtain pdf representations for the joint response-response and response-excitation at different time instants.

{\textbf{Joint response-excitation pdf.} We first formulate the joint response-excitation pdf at two different (arbitrary) time instants. In order to design the joint pdf based on the given marginals of response and excitation, we utilize a bivariate Gaussian copula whose density can be written as follows \cite{Meyer13}:
\small
\begin{align}
\mathcal{C}^{}\left(u,v\right) = \frac{1}{\sqrt{1-c^2}}\exp{\left(\frac{2c \Phi^{-1}\left(u\right)\Phi^{-1}\left(v\right)-c^2\left( \Phi^{-1}\left(u\right)^2 + \Phi^{-1}\left(v\right)^2  \right)}{2(1-c^2)}\right)},
\end{align}
\normalsize
where $u$ and $v$ indicate cumulative distribution functions and  the standard cumulative distribution function is given as the following form:
\begin{align}
\Phi(x) = \frac{1}{\sqrt{2\pi}}\int^{x}_{-\infty}\exp{\left( -\frac{z^2}{2} \right)}dz.
\end{align}
Denoting with $x$ the argument that corresponds to the response at time $t$, with $y$ the argument for the excitation at time $s = t - \tau$, and with $g(y)$ the (zero-mean) marginal pdf for the excitation, we have the expression for the joint response-excitation pdf.
\small
\begin{align}
q(x,y) =&\  f(x) g(y)\mathcal{C}\left(F(x),G(y)\right),\nonumber\\
=&\ f(x) g(y)\frac{1}{\sqrt{1-c^2}}\exp{\left(\frac{2 c \Phi^{-1}\left(F(x)\right)\Phi^{-1}\left(G(y)\right)-c^2\left( \Phi^{-1}\left(F(x)\right)^2 + \Phi^{-1}\left(G(y)\right)^2  \right)}{2(1-c^2)}\right)},
 \label{res_exc_pdf}
\end{align}
\normalsize
where $c$ defines the correlation between the response and the excitation and has values  $-1 \leq c \leq 1$ and $F(x)$ and $G(y)$ are the cumulative distribution functions obtained through the response marginal pdf, $f(x)$, and the excitation marginal pdf, $g(y)$, respectively. Note that the coefficient $c$ depends on the time difference \(\tau=t-s\) of the response and excitation. This dependence will  be recovered through the resolved second-order moments (over time) between the response and excitation.

{\textbf{Joint response-response pdf.} The joint pdf for two different time instants of the response, denoted as $p(x,z)$, is a special case of what has been presented. In order to avoid confusion, a different notation $z$ is used to represent the response  at a different time instant $s = t - \tau$. We have:
\small
\begin{align}
p(x,z) =&\  f(x) f(z)\mathcal{C}\left(F(x),F(z)\right),\nonumber\\
=&\ f(x) f(z)\frac{1}{\sqrt{1-c^2}}\exp{\left(\frac{2 c \Phi^{-1}\left(F(x)\right)\Phi^{-1}\left(F(z)\right)-c^2\left( \Phi^{-1}\left(F(x)\right)^2 + \Phi^{-1}\left(F(z)\right)^2  \right)}{2(1-c^2)}\right)},
\end{align}
\normalsize
where $c$ is a correlation constant (that depends on the time-difference $\tau$). Note that the response  $z$  at the second time instant  follows the same non-Gaussian pdf corresponding to the single time statistics of the response. 

In Figure \ref{figure2}b, we present the above joint pdf (\ref{res_exc_pdf}) with the marginal $f$ (response) having a bimodal structure and the marginal $g$ (excitation) having a Gaussian structure. For $c=0$ we have independence, which essentially expresses the case of very distant two-time statistics, while as we increase $c$ the correlation between the two variables increases referring to the case of small values of $\tau$.

\subsubsection{Induced Non-Gaussian Closures}

\begin{figure}[H]
\centerline{
\begin{minipage}{0.55\hsize}\begin{center}
\includegraphics[width=\hsize]{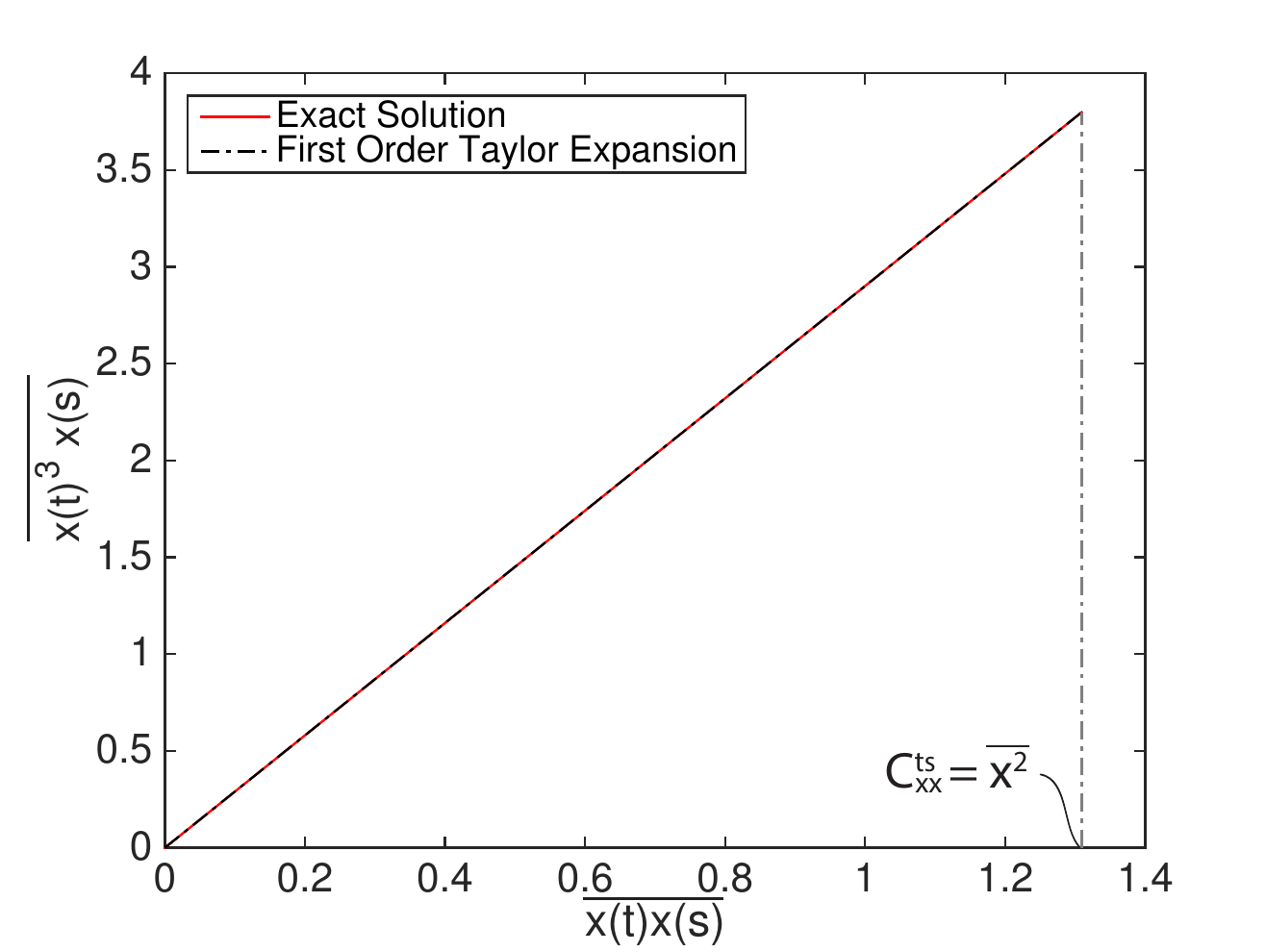}
\end{center}\end{minipage}
}
\caption{The relation between $\overline{x(t)^3x(s)}$ and $\overline{x(t)x(s)}$. Exact relation is illustrated in red curve and approximated relation using non-Gaussian pdf representations is depicted in black curve. }
\label{fig_x3x}
\end{figure}

Using these non-Gaussian pdf representations, we will approximate the fourth order moment terms that show up in the moment equations. We numerically observe that in the context of the pdf representations given above, the relation between $\overline{x(t)^3x(s)}$ and $\overline{x(t)x(s)}$ is essentially linear (see Figure \ref{fig_x3x}). To this end, we choose a closure of the following form for both the response-response and the response-excitation terms:}
\begin{align}
\overline{x(t)^3x(s)} =&\ \rho_{x,x}\ \overline{x(t)x(s)},
\end{align}
where $\rho_{x,x}$ is the closure coefficient for the joint response-response statistics.  The value of $\rho_{x,x}$ is obtained by expanding both $\overline{x(t)^3x(s)}$ and $\overline{x(t)x(s)}$ with respect to $c$  keeping up to the first order terms:\small
\begin{align}
\overline{xz} =&\ \iint xz p(x,z) dx dz = 2\Big\{\int x f(x) \erf^{-1}\left(2 F(x)-1\right)dx \Big\}^2 c +\mathcal{O}(c^2),\\
\overline{x^3z} =&\ \iint x^3z p(x,z) dx dz = 2\Big\{\int x^3 f(x) \erf^{-1}\left(2 F(x)-1\right)dx \Big\}\Big\{\int z f(z) \erf^{-1}\left(2 F(z)-1\right)dz\Big\} c +\mathcal{O}(c^2),
\end{align}
\normalsize
where the error function is given by:
\begin{align}
\erf(x) = \frac{2}{\sqrt{\pi}}\int^{x}_0 e^{-t^2} dt.
\end{align}
Thus, we observe that the assumed copula function in combination with the marginal densities prescribe an explicit dependence between fourth- and second-order moments, expressed through the coefficient:
\begin{align}
\rho_{x,x} =\frac{\int x^3 f(x) \erf^{-1}\left(2 F(x)-1\right)dx}{\int x f(x) \erf^{-1}\left(2 F(x)-1\right)dx} .
\label{closure1}
\end{align}}\textit{We emphasize that this closure coefficient does not depend on the time-difference $\tau$  but only on the single time statistics and in particular }\textit{the energy level
of the system, defined by $\gamma$}. To this end, for any given marginal pdf $f$, we can  analytically find what would be the closure coefficient under the assumptions of the adopted copula function. 

The corresponding coefficient for the joint response-excitation statistics $\rho_{x,y}$ can be similarly obtained through a first order expansion of the moments:
\begin{align}
\rho_{x,y} =\frac{\int x^3 f(x) \erf^{-1}\left(2 F(x)-1\right)dx}{\int x f(x) \erf^{-1}\left(2 F(x)-1\right)dx}
\label{rho_xy}
\end{align}
The closure coefficient $\rho_{x, y}$ has exactly the same form with the closure coefficient $\rho_{x, x}$ and it does not depend on the statistical properties of the excitation nor on the time-difference $\tau$ but only on the energy level $\gamma$. We will refer to equations (\ref{closure1}) and (\ref{rho_xy}) as the \textit{closure constraints}. This will be one of the two sets of constraints that we will include in the minimization procedure for the determination of the solution.

 \subsubsection{Closed Moment Equations}
The next step involves the application of above closure scheme on the derived two-time moment equations. By directly applying the induced closure schemes on equations (\ref{mt_tau1}) and (\ref{mt_tau2}), we have the linear set of moment equations for the second-order statistics:
\begin{align}
\frac{\partial^2}{\partial \tau^2} C_{xy}(\tau) +{\lambda}\frac{\partial}{\partial \tau} C_{xy}(\tau) +({k}_1 +\rho_{x, y}{k}_3)C_{xy}(\tau)=&\ \frac{\partial^2}{\partial \tau^2} C_{yy}(\tau),\label{moment_tau1}\\
\frac{\partial^2}{\partial \tau^2} C_{xx}(\tau) +{\lambda}\frac{\partial}{\partial \tau} C_{xx}(\tau) +({k}_1+\rho_{x, x}{k}_3)C_{xx}(\tau)  =&\ {\frac{\partial^2}{\partial \tau^2} C_{xy}(-\tau)}.\label{moment_tau2}
\end{align}
Using the Wiener-Khinchin theorem, we transform the above equations to the corresponding power spectral density equations:
\begin{align}
\{ (j\omega)^2 + {\lambda} (j\omega) +{k}_1 +\rho_{x, y}{k}_3 \} S_{xy}(\omega)=(j\omega)^2 S_{yy}(\omega) ,\label{spect_1} \\
\{ (j\omega)^2 - {\lambda} (j\omega) +{k}_1 +\rho_{x, x}{k}_3 \} S_{xx}(\omega)=(j\omega)^2 S_{xy}(\omega)  .
\end{align}
These equations allow us to obtain an expression for the power spectral density of the response displacement in terms of the excitation spectrum:
\begin{align}
S_{xx}(\omega)=  \Bigg|\frac{\omega^4}{\{{k}_1 +\rho_{x, y}{k}_3  -\omega^2  +j({\lambda}\omega) \}\{ {k}_1 +\rho_{x, x}{k}_3 -\omega^2  -j({\lambda} \omega) \}}\Bigg| S_{yy}(\omega).
\end{align}
Integration of the above equation will give us the variance of the response: 
\small
\begin{align}
\overline{x^2} =\int^\infty_{0} S_{xx}(\omega) d\omega= \int^\infty_{0} \Bigg|\frac{\omega^4}{\{{k}_1 +\rho_{x, y}{k}_3  -\omega^2  +j({\lambda}\omega) \}\{ {k}_1 +\rho_{x, x}{k}_3 -\omega^2  -j({\lambda} \omega) \}}\Bigg| S_{yy}(\omega) d\omega.
\label{spectral_eqn}
\end{align}
\normalsize
The last equation is the second constraint, the \textit{dynamics constraint}, which expresses the second order dynamics of the system. Our goal is to optimally satisfy  it together with the \textit{closure constraints} defined by equations (\ref{closure1}) and (\ref{rho_xy}).

\subsubsection{Moment Equation Copula Closure (MECC) Method}
 
The last step is the minimization of the two set of constraints, the \textit{closure constraints} and the \textit{dynamics constraint}, which have been expressed in terms of the system response variance $\overline{x^2}$. The minimization will be done in terms of the unknown energy level $\gamma$ and the closure coefficients $\rho_{x, x}$ and $\rho_{x, y}$. 

More specifically, we define the following cost function which incorporates our  constraints:

\small
{
\begin{align}
\mathcal{J}(\gamma,\rho_{x, x},\rho_{x, y}) =&\ \Bigg\{  \overline { x^{ 2 } } -\int _{ 0 }^{ \infty  } |\frac { \omega ^{ 4 }S_{ yy }(\omega ) }{ \{ { k }_{ 1 }+\rho _{ x,y }{ k }_{ 3 }-\omega ^{ 2 }+j({ \lambda  }\omega )\} \{ { k }_{ 1 }+\rho _{ x,x }{ k }_{ 3 }-\omega ^{ 2 }-j({ \lambda  }\omega )\}  } |d\omega  \Bigg\}^{ 2 }\nonumber\\
+&\ \Bigg\{ \rho_{x,x} -\frac{\int x^3 f(x) \erf^{-1}\left(2 F(x)-1\right)dx}{\int x f(x) \erf^{-1}\left(2 F(x)-1\right)dx}   \Bigg\}^{2}+ \Bigg\{ \rho_{x,y} -\frac{\int x^3 f(x) \erf^{-1}\left(2 F(x)-1\right)dx}{\int x f(x) \erf^{-1}\left(2 F(x)-1\right)dx}   \Bigg\}^{2}.
\label{cost_fcn}
\end{align}
}
\normalsize 

Note that in the context of statistical linearization only the first constraint is minimized while the closure coefficient is the one that follows exactly from a Gaussian representation for the pdf. In this context there is no attempt to incorporate in an equal manner the mismatch in the dynamics and the pdf representation. The minimization of this cost function essentially allows mismatch for    the equation (expressed through the dynamic constraint) but also for the pdf representation (expressed through the closure constraints). For linear systems and an adopted Gaussian pdf for the response the above cost function vanishes identically.

\section{Applications to Bistable Energy Harvesters under Correlated Excitation}
We apply the presented Moment Equation Copula Closure (MECC) method to two nonlinear vibrational systems. One is a single-degree-of-freedom (SDOF) bistable oscillator with linear damping that simulates energy harvesting while the other is a similar bistable oscillator coupled with an electromechanical energy harvester. For both applications, it is assumed that the stationary stochastic excitation has a power spectral density given by the Pierson-Moskowitz spectrum, which is typical for excitation created by random water waves:
\begin{align}
S(\omega) = q\  \frac{1}{\omega^5} \exp(-\frac{1}{\omega^4}),
\end{align}
where $q$ controls the intensity of the excitation. 
\subsection{SDOF Bistable Oscillator Excited by Colored Noise}
For the colored noise excitation that we just described, we apply the MECC method. We consider a set of system parameters that correspond to a double well potential. Depending on the intensity of the excitation (which is adjusted by the factor $q$), the response of the bistable system `lives' in three possible regimes. If $q$ is very low, the bistable system is trapped in either of the two wells while if $q$ is very high the energy level is above the homoclinic orbit and the system performs cross-well oscillations. Between these two extreme regimes, the stochastic response exhibits combined features and characteristics of both energy levels and it has a highly nonlinear, multi-frequency character {\cite{Dykman85, Dykman88}}. 

\begin{figure}[H]
\centerline{
\begin{minipage}{\hsize}\begin{center}
\includegraphics[width=\hsize]{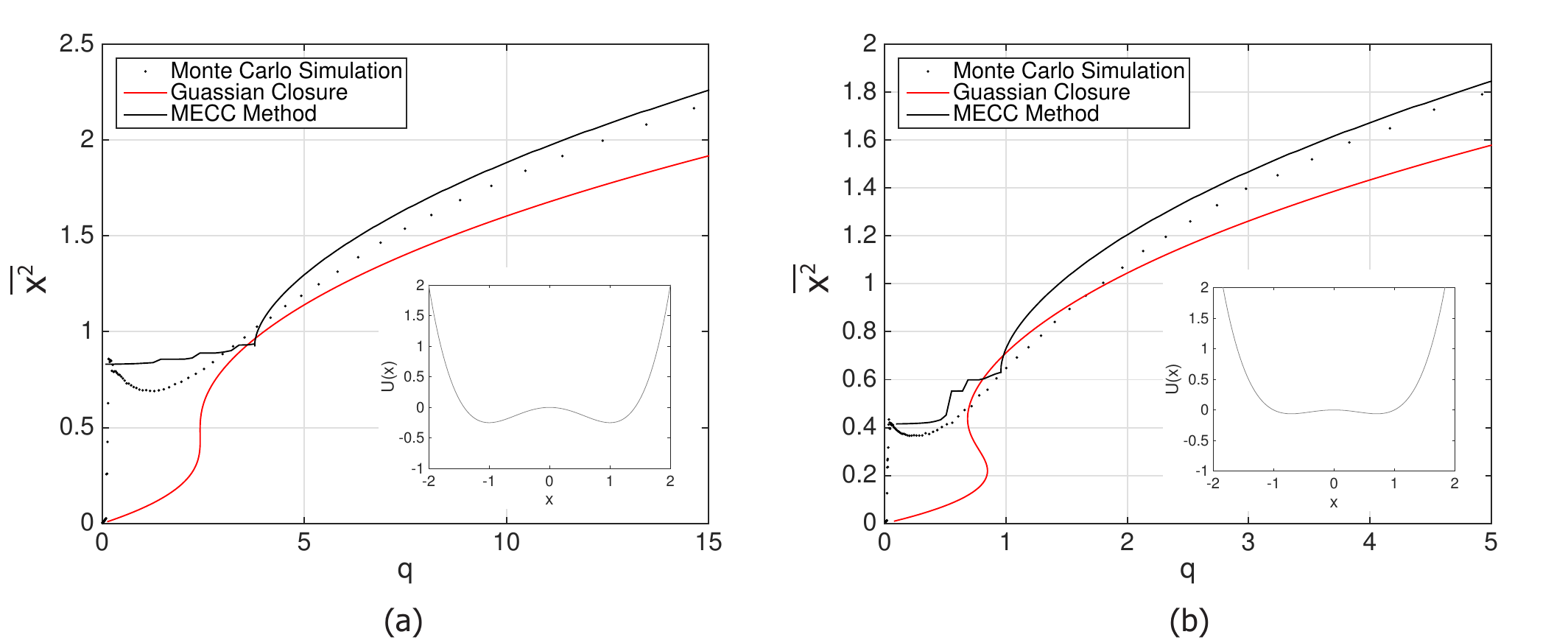}
\end{center}\end{minipage}
}
\caption{Mean square response displacement with respect to the amplification factor of Pierson-Moskowitz spectrum for the bistable system with two different sets of system parameters. (a) ${\lambda}=1$, ${k}_1=-1$, and ${k}_3=1$. (b) ${\lambda}=0.5$, ${k}_1=-0.5$, and ${k}_3=1$.  }
\label{qx2b}
\end{figure}

Despite these challenges, the presented MECC method can inexpensively provide with a very good approximation of the system's statistical characteristics as it is shown in Figure \ref{qx2b}. In particular in Figure \ref{qx2b}, we present the response variance as the intensity of the excitation varies for two sets of the system parameters. We also compare our results with direct Monte-Carlo simulations and with a standard Gaussian closure method {\cite{sobczyk01, Soong93, Grigoriu02}}. 

For the Monte-Carlo simulations the time series for the excitation has been generated as the sum of cosines over a range of frequencies. The amplitudes and the range of frequencies are determined through the power spectrum while the phases are assumed to be random variables which follow a uniform distribution. In the presented examples, the excitation has power spectral density that follows the Pierson-Moskowitz spectrum. Once each ensemble time series for the excitation has been computed, the governing ordinary differential equation is solved using a 4th/5th order Runge-Kutta method. For each realization the system is integrated for a sufficiently long time interval in order to guarantee that the response statistics have converged. For each problem, we generate 100 realizations in order to compute the second-order statistics. However, for the computation of the full joint pdf, a significantly larger number of samples is needed reaching the order of $10^7$. 

We observe that for very large values of $q$ the computed approximation closely follows the Monte-Carlo simulation. On the other hand, the Gaussian closure method systematically underestimates the variance of the response. For lower intensities of the excitation, the exact (Monte-Carlo) variance presents a non-monotonic behavior with respect to $q$ due to the co-existence of the cross- and intra-well oscillations. While the Gaussian closure has very poor performance on capturing this trend, the MECC method can still provide a satisfactory approximation of the dynamics. Note that the non-smooth transition observed in the MECC curve is due to the fact that for very low values of $q$ the minimization of the cost function (\ref{cost_fcn}) does not reach a zero value while this is the case for larger values of $q$. In other words, in the strongly nonlinear regime neither the \textit{dynamics constraint} nor the \textit{closure constraint} is satisfied exactly, yet this optimal solution provides with a good approximation of the system dynamics.

{After we have obtained the unknown parameters $\gamma$, $\rho_{x,x}$ and $\rho_{x,y}$ by minimizing the cost function for each given $q$, we can then compute the covariance functions and the joint pdf in a post-process manner.} More specifically, since a known $\gamma$ corresponds to a specific $\rho_{x,y}$ (equation (\ref{rho_xy})) we can immediately determine $C_{xy}(\tau)$ by taking the inverse Fourier transform of $S_{xy} $ found through equation (\ref{spect_1}). The next step is the numerical integration of the closed moment equation (\ref{moment_tau2}) utilizing the determined value $\rho_{x,x}$ with initial conditions given by
\begin{align}
C_{ xx }(0)=\int { { x }^{ 2 } } f\left( x;\gamma  \right) dx, \quad  and \quad \dot { C } _{ xx }(0)=0,
\end{align}
where the second condition follows from the symmetry properties of $C_{xx}$. Note that we integrate equation (\ref{moment_tau2}) instead of using the inverse Fourier transform as we did for $C_{xy}(\tau)$ so that we can impose the variance found in the last equation by integrating the resulted density for the determined $\gamma$. Using the correlation functions $C_{xx}(\tau)$ and $C_{xy}(\tau)$ we can also determine, for each case, the correlation coefficient $c$ of the copula function for each time-difference $\tau$. The detailed steps are given at the end of this subsection.  

\begin{figure}[H]
\centerline{
\begin{minipage}{\hsize}\begin{center}
\includegraphics[width=\hsize]{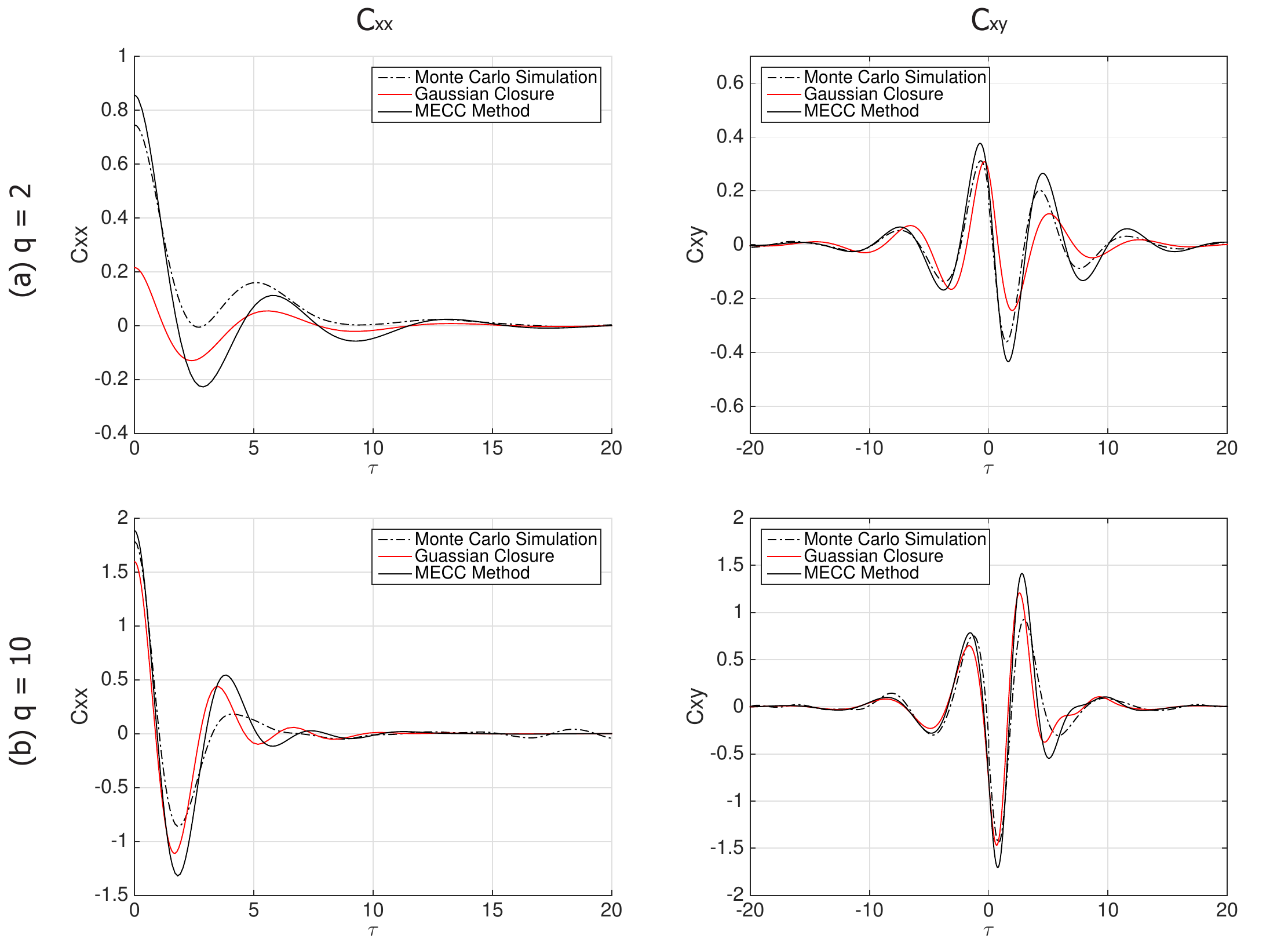}
\end{center}\end{minipage}
}
\caption{Correlation functions $C_{xx}$ and $C_{xy}$ of the bistable system with system parameters ${\lambda}=1$, ${k}_1=-1$, and ${k}_3=1$ subjected to Pierson-Moskowitz spectrum. (a) Amplification factor of $q=2$. (b) Amplification factor of $q=10$. }
\label{corr12}
\end{figure}

The results as well as a comparison with the Gaussian closure method and a direct Monte-Carlo simulation are presented in Figure \ref{corr12}. We can observe that through the proposed approach we are able to satisfactorily approximate the correlation function even close to the non-linear regime $q=2$, where the Gaussian closure method presents important discrepancies.

{
Finally, using the computed parameters $\gamma$ and closure coefficients, $\rho_{x,x}$ and $\rho_{x, y}$, we can also construct the three-dimensional non-Gaussian joint pdf for the response-response-excitation at different time instants. This will be derived based on the three-dimensional Gaussian copula density of the following form:
\small
\begin{align}
\mathcal{C}\left(F(x),F(z),G(y)\right) =\frac{1}{\sqrt{\det{R}}} \exp{\left( -\frac{1}{2}
\left[      
\begin{array}{c}
\Phi^{-1}\left(F(x)\right) \\
\Phi^{-1}\left(F(z)\right)   \\
\Phi^{-1}\left(G(y)\right)   \\
\end{array}
\right]^{T} \cdot \left(R^{-1}-\textbf{I}\right)\cdot
\left[      
\begin{array}{c}
\Phi^{-1}\left(F(x)\right) \\
\Phi^{-1}\left(F(z)\right)   \\
\Phi^{-1}\left(G(y)\right)   \\
\end{array}
\right]
 \right)}.
\end{align}
\normalsize
The three-dimensional non-Gaussian joint pdf for the response-response-excitation at different time instants can be expressed as follows:
\scriptsize
\begin{align}
f_{x(t),x(t+\tau),y(t+\tau)}(x,z,y) =&\ f(x)f(z)g(y)\mathcal{C}\left(F(x),F(z),G(y)\right)\nonumber \\
=&\ f(x)f(z)g(y)\frac{1}{\sqrt{\det{R}}} \exp{\left( -\frac{1}{2}
\left[      
\begin{array}{c}
\Phi^{-1}\left(F(x)\right) \\
\Phi^{-1}\left(F(z)\right)   \\
\Phi^{-1}\left(G(y)\right)   \\
\end{array}
\right]^{T} \cdot \left(R^{-1}-\textbf{I}\right)\cdot
\left[      
\begin{array}{c}
\Phi^{-1}\left(F(x)\right) \\
\Phi^{-1}\left(F(z)\right)   \\
\Phi^{-1}\left(G(y)\right)   \\
\end{array}
\right]
 \right)}.
\end{align}
\normalsize
where $R$ represents the $3 \times\ 3$ correlation matrix with all diagonal elements equal to 1:
\begin{align}
R=\left[      
\begin{array}{ccc}
  1 & c_{xz}  & c_{xy} \\
 c_{xz}   &  1 &  c_{zy} \\
c_{xy}   &  c_{zy} &  1\\
\end{array}
\right].
\end{align}
}The time dependent parameters $c_{xz}$, $c_{xy}$, $c_{zy}$ of the copula function can be found through the resolved moments, by expanding the latter as:
{
\scriptsize
\begin{align}
C_{ xx }(\tau )=&\ \iint { xzf_{x(t),x(t+\tau),y(t+\tau)}(x,z,y)dxdydz } =2\mathcal{F}^2c_{ xz }+\mathcal{O}\left( { c }_{ xz }^{ 2 } \right) ,\\
C_{ xy }(\tau )=&\ \iint { xyf_{x(t),x(t+\tau),y(t+\tau)}(x,z,y)dxdydz }
=2\mathcal{F} \mathcal{G} c_{ xy }+\mathcal{ O }\left( { c }_{ xy }^{ 2 } \right) ,\\
C_{ xy }(0)=&\ \iint { zyf_{x(t),x(t+\tau),y(t+\tau)}(x,z,y)dxdydz } =2\mathcal{F} \mathcal{G} c_{ zy }+\mathcal{ O }\left( { c }^{ 2 }_{ zy } \right).
\end{align}
\normalsize
where,\begin{displaymath}
\mathcal{F}=\int x f(x) \erf^{-1}\left(2 F(x)-1\right)dx \ \ \text{and} \ \ \mathcal{G}=\int x g(x) \erf^{-1}\left(2 G(x)-1\right)dx.
\end{displaymath} If necessary higher order terms may be retained in the Taylor expansion although for the present problem a linear approximation was sufficient. The computed approximation is presented in Figure \ref{3density_sdof} through two dimensional marginals as well as through isosurfaces of the full three-dimensional joint pdf. We compare with direct Monte-Carlo simulations and as we are able to observe, the computed pdf compares favorably with the expensive Monte-Carlo simulation. The joint statistics using the Monte-Carlo approach were computed using $10^7$ number of samples while the computational cost of the MECC method involved the minimization of a three dimensional function.

\noindent

\begin{figure}[H]
\centerline{
\begin{minipage}{\hsize}\begin{center}
\includegraphics[width=\hsize]{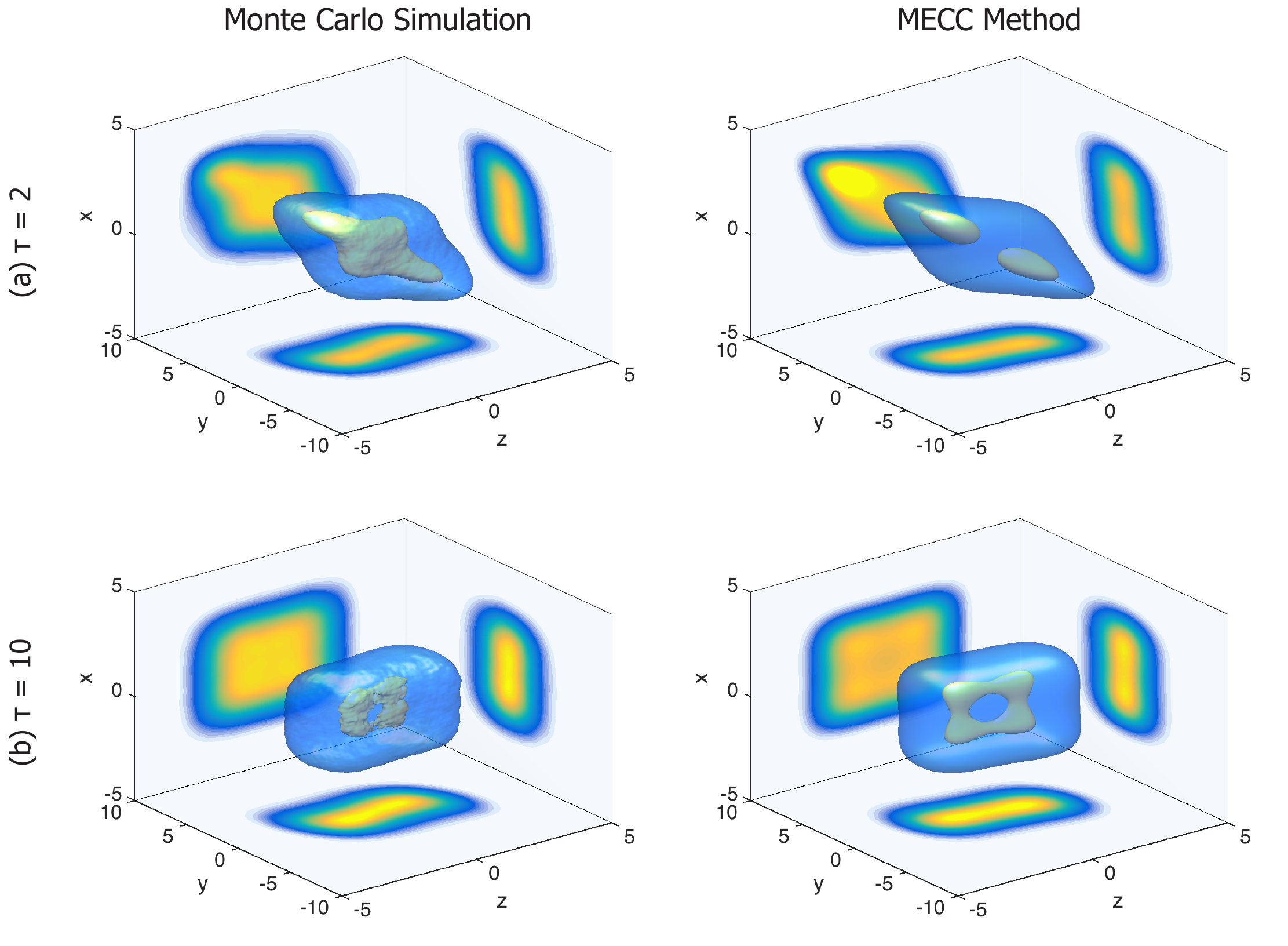}
\end{center}\end{minipage}
}
\caption{Joint pdf $f_{x(t)x(t+\tau)y(t+\tau)}(x,z,y)$ computed using direct Monte-Carlo simulation and the MECC method. The system parameters are given by ${\lambda}=1$, ${k}_1=-1$, and ${k}_3=1$ and the excitation is Gaussian following a Pierson-Moskowitz spectrum with $q=10$. The pdf is presented through two dimensional marginals as well as through isosurfaces. (a) $\tau = 3$. (b) $\tau =10$.}
\label{3density_sdof}
\end{figure}

\subsection{SDOF bistable oscillator coupled to an electromechanical harvester}
In practical configurations, energy harvesting occurs through a linear electromechanical transducer coupled to the nonlinear oscillator \cite{Halvorsen13, Karami12, Masana13}. In this section, we assess how our method performs for a bistable nonlinear SDOF oscillator coupled to a linear electromechanical transducer. The equations of motion in this case take the form:
\begin{align}
\ddot{x} + {\lambda}\dot{x}+{k}_1x+{k}_3x^3+{\alpha} v=&\ \ddot{y}, \label{eom_1}\\
\dot{v} + {\beta} v =&\ {\delta} \dot{x}, \label{eom_2}
\end{align}
 where $x$ is the response displacement, $y$ is a stationary stochastic excitation, $v$ is the voltage across the load, ${\lambda}$ is the normalized damping coefficient, ${k}_1$ and ${k}_3$ are the normalized stiffness coefficients, ${\alpha}$ and ${\delta}$ are the normalized coupling coefficients, and ${\beta}$ is the normalized time coefficient for the electrical system. All the coefficients except ${k}_1$ are positive. Based on the linearity of the second equation, we express the voltage in an integral form:
\begin{align}
v(t) = {\delta} \int^t_0 \dot{x}(\zeta) e^{-{\beta}(t-\zeta)}d\zeta = {\delta}\dot{x}(t)*e^{-{\beta}t} u(t),
\end{align}
where $*$ indicates convolution and $u(t)$ represents the Heaviside step function. We then formulate the second-order moment equations following a similar approach with the previous section.
\begin{align}
\frac{\partial^2}{\partial t^2} \overline{{x}(t)y(s)} +{\lambda} \frac{\partial}{\partial t}\overline{{x}(t)y(s)}+{k}_1\overline{x(t) y(s)}+{k}_3\overline{x(t)^3y(s)} + {\alpha} \overline{v(t) y(s)}= &\ \frac{\partial^2}{\partial t^2}\overline{{y}(t)y(s)}, \label{hdof_mt_tau1} \\
\frac{\partial^2}{\partial t^2} \overline{{x}(t)x(s)} + {\lambda}\frac{\partial}{\partial t} \overline{{x}(t)x(s)}+{k}_1\overline{x(t) x(s)}+{k}_3\overline{x(t)^3x(s)}+ {\alpha} \overline{v(t) x(s)}= &\ \frac{\partial^2}{\partial t^2} \overline{{y}(t)x(s)}, \label{hdof_mt_tau2} \\
\frac{\partial}{\partial t} \overline{v(t)v(s)} +{\beta}\overline{v(t)v(s)} =  &\ {\delta} \frac{\partial}{\partial t} \overline{x(t)v(s)}. \label{hdof_mt_tau3}
\end{align}
In this case, we estimate two additional covariance functions, $\overline{v(t)y(s)}$ and $\overline{v(t)x(s)}$ before applying MECC method: 
\begin{align}
\overline{v(t)y(s)} =&\ {\delta} \int^t_0 \overline{\dot{x}(\zeta)y(s)} e^{-{\beta}(t-\zeta)}d\zeta,\\
=&\ {\delta} \int^t_0\frac{\partial}{\partial \zeta}C_{xy}(\zeta -s) e^{-{\beta}(t-\zeta)}d\zeta,\\
=&\ {\delta} \frac{\partial}{\partial t}C_{xy}(t -s) * e^{-{\beta}t} u(t),\\
=&\ {\delta} \frac{\partial}{\partial \tau}C_{xy}(\tau) * e^{-{\beta}t}u(t),
\end{align}
where $\tau = t-s$ is the time difference of two generic time instants $t$ and $s$. Considering the power spectrum, the Fourier transform of the above gives:
\begin{align}
\mathcal{F}\{\overline{v(t)y(s)} \} =S_{vy}(\omega) = \frac{ j{\delta} \omega}{{\beta}+j\omega}S_{xy}(\omega). 
\end{align}
Similarly, we also obtain for $\overline{v(t)x(s)}$:
\begin{align}
\mathcal{F}\{\overline{v(t)x(s)} \} =S_{vx}(\omega)=\frac{ j{\delta} \omega}{{\beta}+j\omega}S_{xx}(\omega).
\end{align}
By applying the previously described closure scheme on equations (\ref{hdof_mt_tau1}) and (\ref{hdof_mt_tau2}), we have a linear set of moment equations for the second-order statistics:
\begin{align}
\frac{\partial^2}{\partial \tau^2} C_{xy}(\tau) +{\lambda}\frac{\partial}{\partial \tau} C_{xy}(\tau) +({k}_1 +\rho_{x, y}{k}_3)C_{xy}(\tau) +{\alpha} {\delta} \frac{\partial}{\partial \tau}C_{xy}(\tau) * e^{-{\beta}t}u(t)=&\ \frac{\partial^2}{\partial \tau^2} C_{yy}(\tau),\\
\frac{\partial^2}{\partial \tau^2} C_{xx}(\tau) +{\lambda}\frac{\partial}{\partial \tau} C_{xx}(\tau) +({k}_1+\rho_{x, x}{k}_3)C_{xx}(\tau) + {\alpha}{\delta} \frac{\partial}{\partial \tau}C_{xx}(\tau) * e^{-{\beta}t}u(t) =&\ {\frac{\partial^2}{\partial \tau^2} C_{xy}(-\tau)},\\
\frac{\partial}{\partial \tau}C_{vv}(\tau) +{\beta}C_{vv}(\tau)=&\ {\delta}\frac{\partial}{\partial \tau} C_{vx}(-\tau).
\end{align}
Using the Wiener-Khinchin theorem, we transform the above equations to the corresponding power spectral density equations:
\begin{align}
\{ (j\omega)^2 + {\lambda} (j\omega) +{k}_1 +\rho_{x, y}{k}_3+\frac{ j{\alpha}{\delta} \omega}{{\beta}+j\omega} \} S_{xy}(\omega)=&\ (j\omega)^2 S_{yy}(\omega),  \\
\{ (j\omega)^2 - {\lambda} (j\omega) +{k}_1 +\rho_{x, x}{k}_3 -\frac{ j{\alpha}{\delta} \omega}{{\beta}-j\omega}\} S_{xx}(\omega)=&\ (j\omega)^2 S_{xy}(\omega), \\
\{-(j\omega)+{\beta}\}S_{vv}(\omega) =&\ -{\delta} (j\omega)S_{vx}(\omega).
\end{align}
These equations allow us to obtain an expression for the power spectral density of the response displacement and response voltage in terms of the excitation spectrum:
\begin{align}
S_{xx}(\omega)=&\  \Bigg|\frac{\omega^4}{\{{k}_1 +\rho_{x, y}{k}_3  -\omega^2  +j({\lambda}\omega) +\frac{ j{\alpha}{\delta} \omega}{{\beta}+j\omega} \}\{ {k}_1 +\rho_{x, x}{k}_3 -\omega^2  -j({\lambda} \omega) -\frac{ j{\alpha}{\delta} \omega}{{\beta}-j\omega}\}}\Bigg| S_{yy}(\omega),\label{app_spec1}\\
S_{vv}(\omega)
=&\  \Bigg|  \frac{{\delta}^2 \omega^6}{\{{\beta}^2+\omega^2\}\{{k}_1 +\rho_{x, y}{k}_3  -\omega^2  +j({\lambda}\omega) +\frac{ j{\alpha}{\delta} \omega}{{\beta}+j\omega} \}\{ {k}_1 +\rho_{x, x}{k}_3 -\omega^2  -j({\lambda} \omega) -\frac{ j{\alpha}{\delta} \omega}{{\beta}-j\omega}\}}\Bigg| S_{yy}(\omega).
\label{app_spec2}
\end{align}
Integration of the above equation will give us the variance of the response displacement and voltage:
\small
\begin{align}
\overline{x^2}=&\ \int^\infty_{0} \Bigg|\frac{\omega^4}{\{{k}_1 +\rho_{x, y}{k}_3  -\omega^2  +j({\lambda}\omega) +\frac{ j{\alpha}{\delta} \omega}{{\beta}+j\omega} \}\{ {k}_1 +\rho_{x, x}{k}_3 -\omega^2  -j({\lambda} \omega) -\frac{ j{\alpha}{\delta} \omega}{{\beta}-j\omega}\}}\Bigg| S_{yy}(\omega)d\omega,\label{hdof_dispspec}\\
\overline{v^2}=&\ \int^\infty_{0}  \Bigg|  \frac{{\delta}^2 \omega^6}{\{{\beta}^2+\omega^2\}\{{k}_1 +\rho_{x, y}{k}_3  -\omega^2  +j({\lambda}\omega) +\frac{ j{\alpha}{\delta} \omega}{{\beta}+j\omega} \}\{ {k}_1 +\rho_{x, x}{k}_3 -\omega^2  -j({\lambda} \omega) -\frac{ j{\alpha}{\delta} \omega}{{\beta}-j\omega}\}}\Bigg| S_{yy}(\omega)d\omega.
\end{align}
\normalsize
Equation (\ref{hdof_dispspec}) expresses the second order dynamics of the SDOF bistable oscillator coupled with an electromechanical harvester, and is the \textit{dynamics constraint} for this system. We will  minimize it together with the \textit{closure constraints} defined by equations (\ref{closure1}) and (\ref{rho_xy}):
{
\small
\begin{align}
\mathcal{J} (\gamma, \rho_{x,x}, \rho_{x,y})=&\ \Bigg\{ \overline{x^2} - \int^{\infty}_0 \Bigg|\frac{\omega^4 S_{yy}(\omega)}{\{{k}_1 +\rho_{x, y}{k}_3  -\omega^2  +j({\lambda}\omega) +\frac{ j{\alpha}{\delta} \omega}{{\beta}+j\omega} \}\{ {k}_1 +\rho_{x, x}{k}_3 -\omega^2  -j({\lambda} \omega) -\frac{ j{\alpha}{\delta} \omega}{{\beta}-j\omega}\}}\Bigg| d\omega \Bigg\}^2 \nonumber\\
+&\ \Bigg\{ \rho_{x,x} -\frac{\int x^3 f(x) \erf^{-1}\left(2 F(x)-1\right)dx}{\int x f(x) \erf^{-1}\left(2 F(x)-1\right)dx}   \Bigg\}^{ 2 }
+ \Bigg\{ \rho_{x,y} -\frac{\int x^3 f(x) \erf^{-1}\left(2 F(x)-1\right)dx}{\int x f(x) \erf^{-1}\left(2 F(x)-1\right)dx}   \Bigg\}^{ 2 }.
\end{align}
\normalsize
}

In Figure \ref{app_x2b_v2b}, we illustrate the variance of the response displacement and the voltage as the intensity of the excitation varies for two sets of the system parameters. For both sets of system parameters, we observe that for large intensity of the excitation, the MECC method computes the response variances (displacement and voltage) very accurately, while the Gaussian closure method systematically underestimates them. For lower intensities of the excitation, the response displacement variance computed by the Monte-Carlo simulation presents a non-monotonic behavior with respect to $q$. While the Gaussian closure has very poor performance on capturing this trend, the MECC method can still provide a satisfactory approximation of the dynamics. 

\begin{figure}[H]
\centerline{
\begin{minipage}{\hsize}\begin{center}
\includegraphics[width=\hsize]{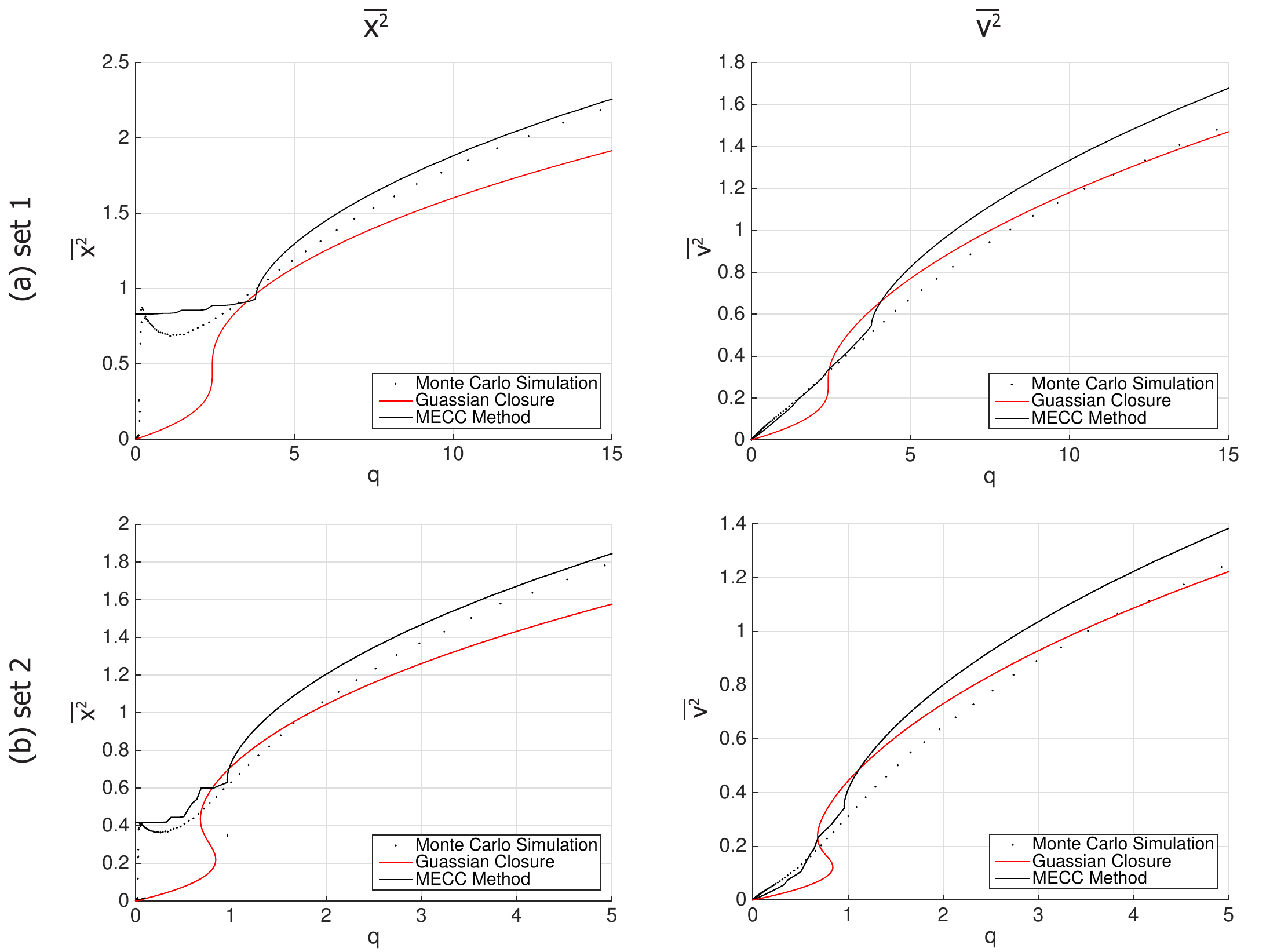}
\end{center}\end{minipage}
}
\caption{Mean square response displacement and mean square response voltage with respect to the amplification factor of Pierson-Moskowitz spectrum for bistable system with two different sets of system parameters. Electromechanical harvester parameters are ${\alpha}=0.01$, ${\beta}=1$, and ${\delta}=1$. {(a) ${\lambda}=1$, ${k}_1=-1$, and ${k}_3=1$.  (b) ${\lambda}=0.5$, ${k}_1=-0.5$, and ${k}_3=1.0$.}}
\label{app_x2b_v2b}
\end{figure}
\noindent
Following similar steps with the previous section, we obtain the covariance functions of the response displacement and voltage and the joint pdf in a post-process manner. The results as well as a comparison with the Gaussian closure method and the Monte-Carlo simulation are illustrated in Figure \ref{hdof_cov}. We can observe that through the proposed approach we are able to satisfactorily approximate the correlation function even close to the non-linear regime $q = 2$, where the Gaussian closure method presents important discrepancies. In Figure \ref{3density_hdof}, we illustrate two dimensional marginal pdfs as well as isosurfaces of the full three-dimensional joint pdf. We compare with direct Monte-Carlo simulations and as we are able to observe, the computed pdf closely approximates the expensive Monte-Carlo simulation in statistical regimes which are far from Gaussian.

\begin{figure}[H]
\centerline{
\begin{minipage}{\hsize}\begin{center}
\includegraphics[width=\hsize]{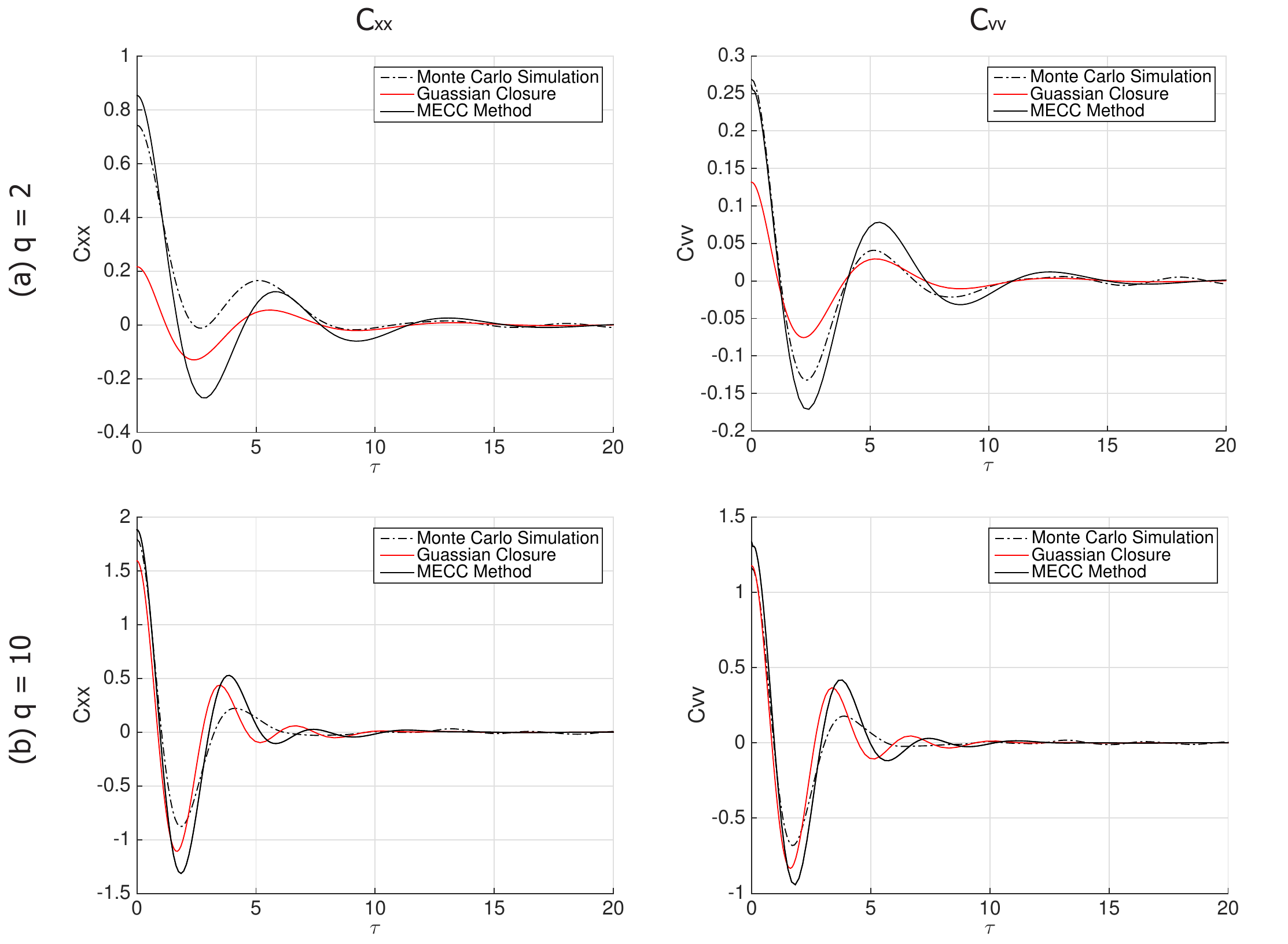}
\end{center}\end{minipage}
}
\caption{Correlation functions $C_{xx}$ and $C_{vv}$  of the bistable system with ${\lambda}=1$, ${k}_1=-1$, and ${k}_3=1$ subjected to Pierson-Moskowitz spectrum. Electromechanical harvester parameters are ${\alpha}=0.01$, ${\beta}=1$, and ${\delta}=1$. (a) Amplification factor of $q=2$. (b) Amplification factor of $q=10$. }
\label{hdof_cov}
\end{figure}

\begin{figure}[H]
\centerline{
\begin{minipage}{\hsize}\begin{center}
\includegraphics[width=\hsize]{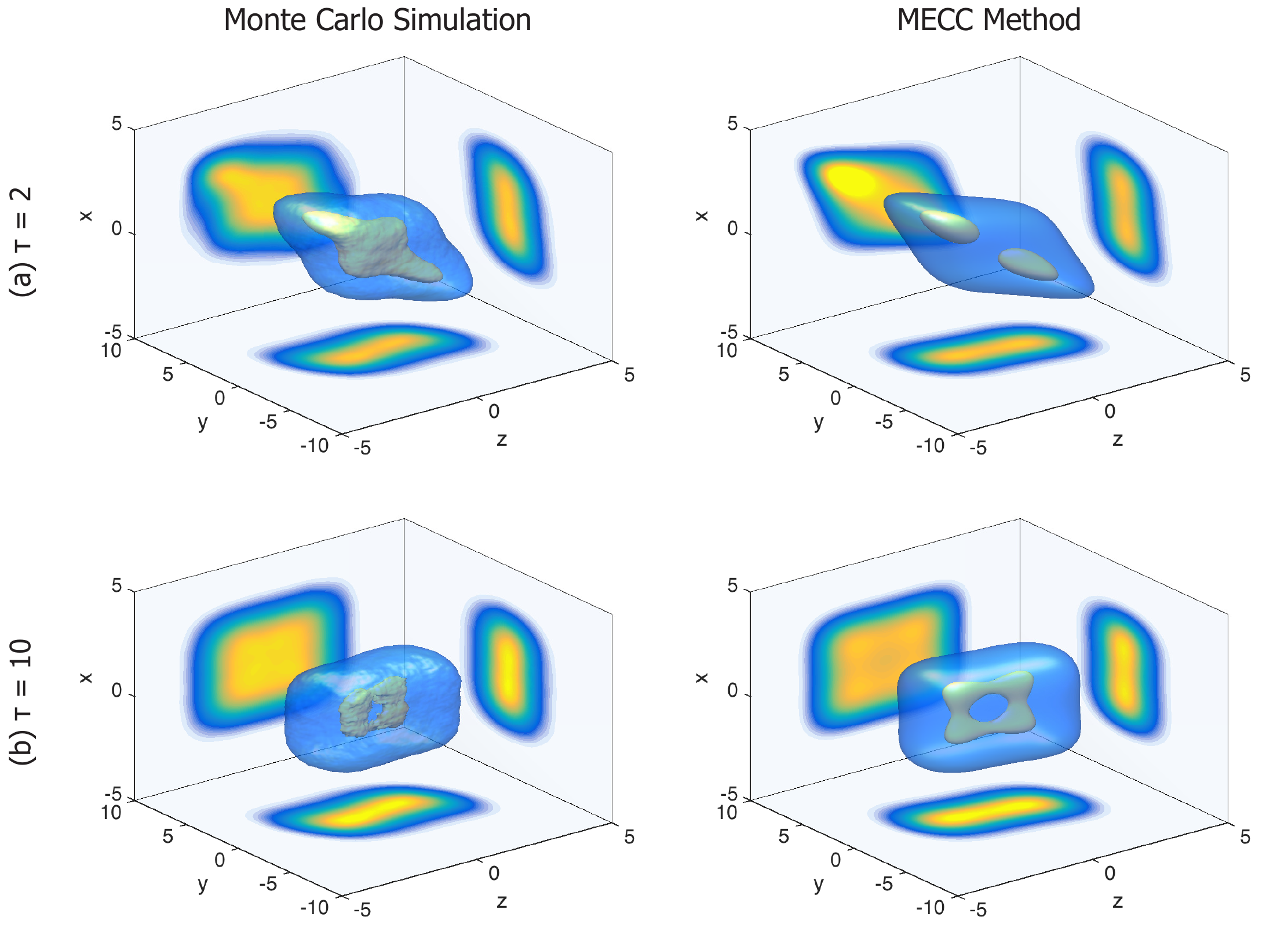}
\end{center}\end{minipage}
}
\caption{Joint pdf $f_{x(t)x(t+\tau)y(t+\tau)}(x,z,y)$ computed using direct Monte-Carlo simulation and the MECC method. The system parameters are given by ${\lambda}=1$, ${k}_1=-1$, and ${k}_3=1$ under Pierson-Moskowitz spectrum $q=10$. Electromechanical harvester parameters are ${\alpha}=0.01$, ${\beta}=1$, and ${\delta}=1$. The pdf is presented through two dimensional marginals as well as through isosurfaces. (a) $\tau = 3$. (b) $\tau =10$.}
\label{3density_hdof}
\end{figure}
Finally in Figure \ref{3d_cmp}, we demonstrate how the proposed MECC method can be used to study robustness over variations of the excitation parameters. In particular, we present the mean square response displacement and response voltage estimated for various amplification factors $q$ and frequency-varied excitation spectra:
\begin{align}
S_p(\omega) = S(\omega-\omega_0),
\end{align}
where $\omega_0$ is the perturbation frequency. The comparison with direct Monte-Carlo simulation indicates the effectiveness of the presented method to capture accurately the response characteristics  over a wide range of input parameters.  

\begin{figure}[H]
\centerline{
\begin{minipage}{\hsize}\begin{center}
\includegraphics[width=\hsize]{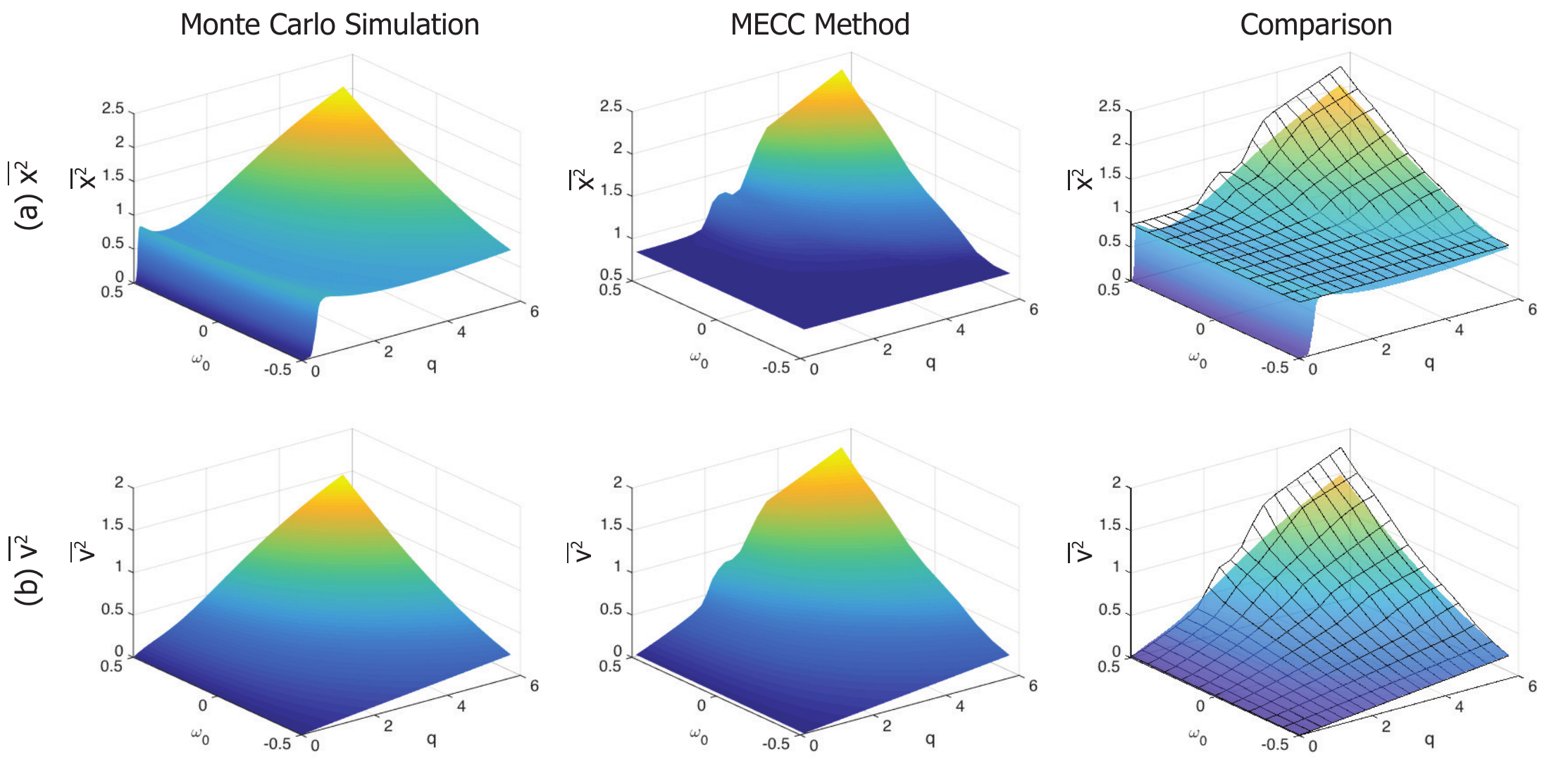}
\end{center}\end{minipage}
}
\caption{Performance comparison (mean square response displacement (a) and voltage (b)) between Monte-Carlo simulations (100 realizations) and MECC method. Results are shown in terms of the amplification factor $q$ and the perturbation frequency $\omega_0$ of the excitation spectrum (Pierson-Moskowitz) for the bistable system with ${\lambda}=1$, ${k}_1=-1$, and ${k}_3=1$. The electromechanical harvester parameters are ${\alpha}=0.01$, ${\beta}=1$, and ${\delta}=1$.}
\label{3d_cmp}
\end{figure}

\section{Conclusions}
We have considered the problem of determining the non-Gaussian steady state statistical structure of bistable nonlinear vibrational systems subjected to colored noise excitation. We first derived moment equations that describe the dynamics governing the two-time statistics. We then combined those with a non-Gaussian pdf representation for the joint response-response and joint response-excitation statistics. This representation has i) single time statistical structure consistent with the analytical solutions of the Fokker-Planck equation, and ii) two-time statistical structure that follows from the adoption of a Gaussian copula function. The pdf representation takes the form of \textit{closure constraints} while the moment equations have the form of a \textit{dynamics constraint}. We  formulated the two sets of constraints as a low-dimensional minimization problem with respect to the unknown parameters of the representation. The minimization of both the \textit{dynamics constraint} and the \textit{closure constraints} imposes an interplay between these two factors. 

We then applied the presented method to two nonlinear oscillators in the context of vibration energy harvesting. One is a single degree of freedom (SDOF) bistable oscillator with linear damping while the other is a same SDOF bistable oscillator coupled with an electromechanical energy harvester. For both applications, it was assumed that the stationary stochastic excitation has a power spectral density given by the Pierson-Moskowitz spectrum. We have shown that the presented method can provide a very good approximation of second order statistics of the system, when compared with direct Monte-Carlo simulations, even in essentially nonlinear regimes, where Gaussian closure techniques fail completely to capture the dynamics. In addition, we can compute the full (non-Gaussian) probabilistic structure of the solution in a post-process manner. We emphasize that the computational cost associated with the new method is considerably smaller compared with methods that evolve the pdf of the solution since MECC method relies on the minimization of a function with a few unknown variables. 

These results indicate that the new method can be a very good candidate when it comes to the calculation of the stochastic response for vibrational system with complex potentials as it is required in parameter optimization or selection. Future endeavors include the application of the presented approach in higher dimensional contexts involving nonlinear energy harvesters and passive protection of structures as well as on the development/optimization of structural configurations able to operate effectively under intermittent loads \cite{Mohamad15}.

\section*{Acknowledgments}

We would like to thank the anonymous referees for helpful suggestions and in particular one of them for suggesting the use of copula functions, a change that significantly improved the original version of the manuscript. We would also like to acknowledge the support from the Samsung Scholarship Program as well as the MIT Energy Initiative for support under the grant `Nonlinear Energy Harvesting From Broad-Band Vibrational Sources By Mimicking Turbulent Energy Transfer Mechanisms'. T.P.S. is also grateful to the American Bureau of Shipping for support under a Career Development Chair.

\bibliographystyle{ieeetr}
\bibliography{ngc}

\begin{thebibliography}{10}

\bibitem{Grigoriu02}
M.~Grigoriu, {\em Stochastic calculus: applications in science and
  engineering}.
\newblock Springer, 2002.

\bibitem{Stratonovich67}
R.~Stratonovich, {\em Topics in the theory of random noise}, vol.~2.
\newblock CRC Press, 1967.

\bibitem{sobczyk01}
K.~Sobczyk, {\em Stochastic differential equations: with applications to
  physics and engineering}, vol.~40.
\newblock Springer, 2001.

\bibitem{Soong93}
T.~Soong and M.~Grigoriu, ``Random vibration of mechanical and structural
  systems,'' {\em NASA STI/Recon Technical Report A}, vol.~93, p.~14690, 1993.

\bibitem{Naess12}
A.~Naess and T.~Moan, {\em Stochastic dynamics of marine structures}.
\newblock Cambridge University Press, 2012.

\bibitem{To11}
C.~To, {\em Nonlinear random vibration: Analytical techniques and
  applications}.
\newblock CRC Press, 2011.

\bibitem{Wojtkiewicz99}
S.~Wojtkiewicz, E.~Johnson, L.~Bergman, M.~Grigoriu, and B.~Spencer~Jr,
  ``Response of stochastic dynamical systems driven by additive gaussian and
  poisson white noise: Solution of a forward generalized kolmogorov equation by
  a spectral finite difference method,'' {\em Computer methods in applied
  mechanics and engineering}, vol.~168, no.~1, pp.~73--89, 1999.

\bibitem{Dunne97}
J.~Dunne and M.~Ghanbari, ``Extreme-value prediction for non-linear stochastic
  oscillators via numerical solutions of the stationary fpk equation,'' {\em
  Journal of Sound and Vibration}, vol.~206, no.~5, pp.~697--724, 1997.

\bibitem{Paola02}
M.~Di~Paola and A.~Sofi, ``Approximate solution of the
  fokker--planck--kolmogorov equation,'' {\em Probabilistic Engineering
  Mechanics}, vol.~17, no.~4, pp.~369--384, 2002.

\bibitem{Wehner83}
M.~F. Wehner and W.~Wolfer, ``Numerical evaluation of path-integral solutions
  to fokker-planck equations,'' {\em Physical Review A}, vol.~27, no.~5,
  p.~2663, 1983.

\bibitem{Naess93}
A.~Naess and J.~Johnsen, ``Response statistics of nonlinear, compliant offshore
  structures by the path integral solution method,'' {\em Probabilistic
  Engineering Mechanics}, vol.~8, no.~2, pp.~91--106, 1993.

\bibitem{Paola08}
M.~Di~Paola and R.~Santoro, ``Path integral solution for non-linear system
  enforced by poisson white noise,'' {\em Probabilistic Engineering Mechanics},
  vol.~23, no.~2, pp.~164--169, 2008.

\bibitem{Kougioumtzoglou12}
I.~Kougioumtzoglou and P.~Spanos, ``An analytical wiener path integral
  technique for non-stationary response determination of nonlinear
  oscillators,'' {\em Probabilistic Engineering Mechanics}, vol.~28,
  pp.~125--131, 2012.

\bibitem{Kougioumtzoglou14}
I.~A. Kougioumtzoglou and P.~D. Spanos, ``Nonstationary stochastic response
  determination of nonlinear systems: A wiener path integral formalism,'' {\em
  Journal of Engineering Mechanics}, vol.~140, no.~9, p.~04014064, 2014.

\bibitem{Matteo14}
A.~Di~Matteo, I.~A. Kougioumtzoglou, A.~Pirrotta, P.~D. Spanos, and
  M.~Di~Paola, ``Stochastic response determination of nonlinear oscillators
  with fractional derivatives elements via the wiener path integral,'' {\em
  Probabilistic Engineering Mechanics}, vol.~38, pp.~127--135, 2014.

\bibitem{Sapsis08}
T.~P. Sapsis and G.~A. Athanassoulis, ``New partial differential equations
  governing the joint, response--excitation, probability distributions of
  nonlinear systems, under general stochastic excitation,'' {\em Probabilistic
  Engineering Mechanics}, vol.~23, no.~2, pp.~289--306, 2008.

\bibitem{Venturi12}
D.~Venturi, T.~P. Sapsis, H.~Cho, and G.~E. Karniadakis, ``A computable
  evolution equation for the joint response-excitation probability density
  function of stochastic dynamical systems,'' {\em Proceedings of the Royal
  Society A: Mathematical, Physical and Engineering Science}, vol.~468,
  no.~2139, pp.~759--783, 2012.

\bibitem{Cho13}
H.~Cho, D.~Venturi, and G.~E. Karniadakis, ``Adaptive discontinuous galerkin
  method for response-excitation pdf equations,'' {\em SIAM Journal on
  Scientific Computing}, vol.~35, no.~4, pp.~B890--B911, 2013.

\bibitem{Caughey59}
T.~Caughey, ``Response of a nonlinear string to random loading,'' {\em Journal
  of Applied Mechanics}, vol.~26, no.~3, pp.~341--344, 1959.

\bibitem{Caughey63}
T.~Caughey, ``Equivalent linearization techniques,'' {\em The Journal of the
  Acoustical Society of America}, vol.~35, no.~11, pp.~1706--1711, 1963.

\bibitem{Kazakov54}
I.~Kazakov, ``An approximate method for the statistical investigation of
  nonlinear systems,'' {\em Trudy VVIA im Prof. NE Zhukovskogo}, vol.~394,
  pp.~1--52, 1954.

\bibitem{Roberts03}
J.~Roberts and P.~Spanos, {\em Random vibration and statistical linearization}.
\newblock Courier Dover Publications, 2003.

\bibitem{Socha08}
L.~Socha, {\em Linearization methods for stochastic dynamic systems}, vol.~730.
\newblock Springer, 2008.

\bibitem{Crandall05}
S.~H. Crandall, ``On using non-gaussian distributions to perform statistical
  linearization,'' vol.~39, p.~1395, 2004.

\bibitem{Sancho70}
N.~Sancho, ``Technique for finding the moment equations of a nonlinear
  stochastic system,'' {\em Journal of Mathematical Physics}, vol.~11, no.~3,
  pp.~771--774, 1970.

\bibitem{Bover78}
D.~Bover, ``Moment equation methods for nonlinear stochastic systems,'' {\em
  Journal of Mathematical Analysis and Applications}, vol.~65, no.~2,
  pp.~306--320, 1978.

\bibitem{Beran94}
J.~Beran, {\em Statistics for long-memory processes}, vol.~61.
\newblock CRC Press, 1994.

\bibitem{Iyengar78}
R.~Iyengar and P.~Dash, ``Study of the random vibration of nonlinear systems by
  the gaussian closure technique,'' {\em Journal of Applied Mechanics},
  vol.~45, no.~2, pp.~393--399, 1978.

\bibitem{Crandall80}
S.~Crandall, ``Non-gaussian closure for random vibration of non-linear
  oscillators,'' {\em International Journal of Non-Linear Mechanics}, vol.~15,
  no.~4, pp.~303--313, 1980.

\bibitem{Crandall85}
S.~Crandall, ``Non-gaussian closure techniques for stationary random
  vibration,'' {\em International journal of non-linear mechanics}, vol.~20,
  no.~1, pp.~1--8, 1985.

\bibitem{Liu88}
Q.~Liu and H.~Davies, ``Application of non-gaussian closure to the
  nonstationary response of a duffing oscillator,'' {\em International journal
  of non-linear mechanics}, vol.~23, no.~3, pp.~241--250, 1988.

\bibitem{Wu84}
W.~Wu and Y.~Lin, ``Cumulant-neglect closure for non-linear oscillators under
  random parametric and external excitations,'' {\em International Journal of
  Non-Linear Mechanics}, vol.~19, no.~4, pp.~349--362, 1984.

\bibitem{Ibrahim85}
R.~Ibrahim, A.~Soundararajan, and H.~Heo, ``Stochastic response of nonlinear
  dynamic systems based on a non-gaussian closure,'' {\em Journal of applied
  mechanics}, vol.~52, no.~4, pp.~965--970, 1985.

\bibitem{Grigoriu91}
M.~Grigoriu, ``A consistent closure method for non-linear random vibration,''
  {\em International journal of non-linear mechanics}, vol.~26, no.~6,
  pp.~857--866, 1991.

\bibitem{Hasofer95}
A.~Hasofer and M.~Grigoriu, ``A new perspective on the moment closure method,''
  {\em Journal of applied mechanics}, vol.~62, no.~2, pp.~527--532, 1995.

\bibitem{Wojtkiewicz96}
S.~Wojtkiewicz, B.~Spencer~Jr, and L.~Bergman, ``On the cumulant-neglect
  closure method in stochastic dynamics,'' {\em International journal of
  non-linear mechanics}, vol.~31, no.~5, pp.~657--684, 1996.

\bibitem{Grigoriu99}
M.~Grigoriu, ``Moment closure by monte carlo simulation and moment sensitivity
  factors,'' {\em International journal of non-linear mechanics}, vol.~34,
  no.~4, pp.~739--748, 1999.

\bibitem{Noori87}
M.~Noori, A.~Saffar, and H.~Davoodi, ``A comparison between non-gaussian
  closure and statistical linearization techniques for random vibration of a
  nonlinear oscillator,'' {\em Computers \& structures}, vol.~26, no.~6,
  pp.~925--931, 1987.

\bibitem{Green12}
P.~Green, K.~Worden, K.~Atallah, and N.~Sims, ``The benefits of duffing-type
  nonlinearities and electrical optimisation of a mono-stable energy harvester
  under white gaussian excitations,'' {\em Journal of Sound and Vibration},
  vol.~331, no.~20, pp.~4504--4517, 2012.

\bibitem{harne13}
R.~Harne and K.~Wang, ``A review of the recent research on vibration energy
  harvesting via bistable systems,'' {\em Smart Materials and Structures},
  vol.~22, no.~2, p.~023001, 2013.

\bibitem{Daqaq11}
M.~Daqaq, ``Transduction of a bistable inductive generator driven by white and
  exponentially correlated gaussian noise,'' {\em Journal of Sound and
  Vibration}, vol.~330, no.~11, pp.~2554--2564, 2011.

\bibitem{Halvorsen13}
E.~Halvorsen, ``Fundamental issues in nonlinear wideband-vibration energy
  harvesting,'' {\em Physical Review E}, vol.~87, no.~4, p.~042129, 2013.

\bibitem{Green13}
P.~Green, E.~Papatheou, and N.~Sims, ``Energy harvesting from human motion and
  bridge vibrations: An evaluation of current nonlinear energy harvesting
  solutions,'' {\em Journal of Intelligent Material Systems and Structures},
  2013.

\bibitem{He14}
Q.~He and M.~Daqaq, ``New insights into utilizing bi-stability for energy
  harvesting under white noise,'' {\em Journal of Vibration and Acoustics},
  2014.

\bibitem{Mann09}
B.~Mann and N.~Sims, ``Energy harvesting from the nonlinear oscillations of
  magnetic levitation,'' {\em Journal of Sound and Vibration}, vol.~319, no.~1,
  pp.~515--530, 2009.

\bibitem{Barton10}
D.~Barton, S.~Burrow, and L.~Clare, ``Energy harvesting from vibrations with a
  nonlinear oscillator,'' {\em Journal of Vibration and Acoustics}, vol.~132,
  no.~2, p.~021009, 2010.

\bibitem{Joo14}
H.~K. Joo and T.~P. Sapsis, ``Performance measures for single-degree-of-freedom
  energy harvesters under stochastic excitation,'' {\em Journal of Sound and
  Vibration}, vol.~333, no.~19, pp.~4695--4710, 2014.

\bibitem{Kluger15}
J.~M. Kluger, T.~P. Sapsis, and A.~H. Slocum, ``Robust energy harvesting from
  walking vibrations by means of nonlinear cantilever beams,'' {\em Journal of
  Sound and Vibration}, vol.~341, pp.~174--194, 2015.

\bibitem{Athanassoulis13}
G.~Athanassoulis, I.~Tsantili, and Z.~Kapelonis, ``Two-time,
  response-excitation moment equations for a cubic half-oscillator under
  gaussian and cubic-gaussian colored excitation. part 1: The monostable
  case,'' {\em arXiv preprint arXiv:1304.2195}, 2013.

\bibitem{Daqaq12}
M.~Daqaq, ``On intentional introduction of stiffness nonlinearities for energy
  harvesting under white gaussian excitations,'' {\em Nonlinear Dynamics},
  vol.~69, no.~3, pp.~1063--1079, 2012.

\bibitem{Gammaitoni09}
L.~Gammaitoni, I.~Neri, and H.~Vocca, ``Nonlinear oscillators for vibration
  energy harvesting,'' {\em Applied Physics Letters}, vol.~94, no.~16,
  p.~164102, 2009.

\bibitem{Ferrari10}
M.~Ferrari, V.~Ferrari, M.~Guizzetti, B.~And{\`o}, S.~Baglio, and C.~Trigona,
  ``Improved energy harvesting from wideband vibrations by nonlinear
  piezoelectric converters,'' {\em Sensors and Actuators A: Physical},
  vol.~162, no.~2, pp.~425--431, 2010.

\bibitem{Isserlis18}
L.~Isserlis, ``On a formula for the product-moment coefficient of any order of
  a normal frequency distribution in any number of variables,'' {\em
  Biometrika}, pp.~134--139, 1918.

\bibitem{Soize94}
C.~Soize, {\em The Fokker-Planck equation for stochastic dynamical systems and
  its explicit steady state solutions}, vol.~17.
\newblock World Scientific, 1994.

\bibitem{Nelsen07}
R.~B. Nelsen, {\em An introduction to copulas}.
\newblock Springer Science \& Business Media, 2007.

\bibitem{Meyer13}
C.~Meyer, ``The bivariate normal copula,'' {\em Communications in
  Statistics-Theory and Methods}, vol.~42, no.~13, pp.~2402--2422, 2013.

\bibitem{Qu12}
L.~Qu and W.~Yin, ``Copula density estimation by total variation penalized
  likelihood with linear equality constraints,'' {\em Computational Statistics
  \& Data Analysis}, vol.~56, no.~2, pp.~384--398, 2012.

\bibitem{Dykman85}
M.~Dykman, S.~Soskin, and M.~Krivoglaz, ``Spectral distribution of a nonlinear
  oscillator performing brownian motion in a double-well potential,'' {\em
  Physica A: Statistical Mechanics and its Applications}, vol.~133, no.~1,
  pp.~53--73, 1985.

\bibitem{Dykman88}
M.~Dykman, R.~Mannella, R.~McClintock, F.~Moss, and S.~Soskin, ``Spectral
  density of fluctuations of a double-well duffing oscillator driven by white
  noise,'' {\em Physical Review A}, vol.~37, no.~4, p.~1303, 1988.

\bibitem{Karami12}
M.~Karami and D.~Inman, ``Powering pacemakers from heartbeat vibrations using
  linear and nonlinear energy harvesters,'' {\em Applied Physics Letters},
  vol.~100, no.~4, p.~042901, 2012.

\bibitem{Masana13}
R.~Masana and M.~Daqaq, ``Response of duffing-type harvesters to band-limited
  noise,'' {\em Journal of Sound and Vibration}, vol.~332, no.~25,
  pp.~6755--6767, 2013.

\bibitem{Mohamad15}
M.~Mohamad and T.~Sapsis, ``Probabilistic description of extreme events in
  intermittently unstable dynamical systems excited by correlated stochastic
  processes,'' {\em SIAM/ASA Journal on Uncertainty Quantification}, vol.~3,
  no.~1, pp.~709--736, 2015.

\end{thebibliography}

\end{document}